\documentclass{aa}  

\usepackage{graphicx}
\usepackage{txfonts}
\usepackage{natbib}
\bibpunct{(}{)}{;}{a}{}{,} 

\begin{document}

\title{The KSW estimator for oscillations in the CMB bispectrum}

\author{Moritz M\"{u}nchmeyer\inst{\ref{inst1},\ref{inst2},\ref{inst3}}
\and Fran\c{c}ois Bouchet\inst{\ref{inst1},\ref{inst2},\ref{inst3}}
\and Mark G. Jackson\inst{\ref{inst1},\ref{inst2},\ref{inst6},\ref{inst7}}
\and Benjamin Wandelt\inst{\ref{inst1},\ref{inst2},\ref{inst3},\ref{inst4},\ref{inst5}}}

\institute{
Sorbonne Universit\'{e}s, UPMC Univ Paris 06, UMR7095, Institut d'Astrophysique de Paris, F-75014, Paris, France \label{inst1}
\and CNRS, UMR7095, Institut d'Astrophysique de Paris, F-75014, Paris, France \label{inst2}
\and Lagrange Institute (ILP) 98 bis, boulevard Arago 75014 Paris France \label{inst3}
\and Department of Physics, University of Illinois at Urbana-Champaign, Urbana, IL 61801 \label{inst4}
\and Department of Astronomy, University of Illinois at Urbana-Champaign, Urbana, IL 61801 \label{inst5}
\and Paris Centre for Cosmological Physics and Laboratoire AstroParticule et Cosmologie, \\Universit\'{e} Paris 7-Denis Diderot, Paris, France \label{inst6}
\and African Institute for Mathematical Sciences, Muizenberg, 7945, South Africa \label{inst7}}

\date{Received 08 September 2014}

\abstract{Oscillating shapes of the primordial bispectrum are present in many inflationary models. The Planck experiment has recently published measurements of oscillating shapes, which were, however, limited to the efficient frequency range of the used analysis method. Here, we study the Komatsu Spergel Wandelt (KSW) estimator for oscillations in the cosmic microwave background (CMB) bispectrum, that examines arbitrary oscillation frequencies for separable oscillating bispectrum shapes. We study the precision with which amplitude, phase, and frequency can be determined with our estimator. An examination of the three-point function in real space gives further insight into the estimator.}

\maketitle

\section{Introduction}

The cosmic microwave background (CMB) provides the most direct experimental access to the statistics of the primordial curvature perturbations. In standard single-field, slow roll inflation, these perturbations are Gaussian to a very good approximation~(\cite{Maldacena:2002vr}). However, more complicated models of inflation often predict detectable amounts of non-Gaussianities of various shapes. In particular, the primordial bispectrum $B(k_1,k_2,k_3)$ arising from the three-point correlations of the curvature field $\left< \Phi \Phi \Phi \right>$ is a sensitive probe to discriminate among models of inflation (see e.g. the reviews~\cite{Liguori:2010hx,Komatsu:2010hc,Yadav:2010fz}). High precision measurements of the CMB recently provided by the Planck experiment~(\cite{Ade:2013ydc}) have been consistent with Gaussianity and set stronger limits on primordial non-Gaussianities. However, the availability of such limits depend on the specific shape of the bispectra under consideration. 

A class of bispectra that has attracted attention in recent years are oscillating shapes. Such bispectrum oscillations can arise in a variety of theoretical models. The authors of~\cite{Chen:2008wn} calculated the primordial bispectrum in the presence of features in the inflaton potential of standard single field inflation. They provided two analytical bispectrum shapes that approximate their results. The \textit{feature model} oscillates linearly with the scale and is induced by sharp features in the inflaton potential. The \textit{resonance model} includes oscillations with the logarithm of the scale and is induced by periodic features in the inflaton potential. More recently, the authors of~\cite{Bartolo:2010bj} used the effective field theory of inflation to examine the influence of sharp features in the inflaton potential, which also provide oscillating bispectrum solutions. Periodical modulations of the inflaton potential appear, for example, in axion monodromy inflation models~(\cite{Flauger:2009ab,Flauger:2010ja}). Certain bispectrum shapes motivated by Non-Bunch-Davis vacua also include oscillations (e.g.~\cite{Chen:2006nt,Meerburg:2009ys}), as do cascade inflation models~(\cite{Ashoorioon:2006wc}). A transient reduction in the speed of sound also leads to oscillations ~(\cite{Achucarro:2013cva,Achucarro:2014msa}). Oscillations in the bispectrum are usually accompanied by oscillations in the primordial power spectrum. Recent searches for power spectrum oscillation were presented in~\cite{Pahud:2008ae,Peiris:2013opa,Ade:2013uln,Easther:2013kla,Meerburg:2013cla,Meerburg:2013dla}, but no statistically significant result has been found. Combining power spectrum and bispectrum measurements can improve the sensitivity.

In the present work, we focus on the feature model shape, since it is simple and approximates some more complicated oscillating shapes. In particular, the feature model has the important property of separability. The Planck paper on non-Gaussianities~(\cite{Ade:2013ydc}) already included a targeted search for the feature and resonant shapes with the modal expansion method. In this methodology, the bispectrum under consideration is expanded into a basis of separable shapes~(\cite{Fergusson:2009nv}), which allows an efficient numerical estimation by the Komatsu Spergel Wandelt (KSW) estimator~(\cite{Komatsu:2003iq}). However, the separable basis functions used by Planck did not allow to represent high frequency oscillations, limiting the frequency range, which could be searched for oscillations. The situation can be improved by using a set of oscillating basis functions for the modal expansion~(\cite{Meerburg:2010ks}). In this work, we follow a more direct approach by writing the feature model bispectrum in separable form, which makes it possible to search for oscillations of arbitrary frequency, which are only limited by the resolution of the maps. The gain in computational time with respect to the modal expansion is of the order of the number of modes that would be necessary to approximate the shape. We study the properties of the estimator in detail, including the precision with which phase and frequency of the oscillation can be determined. We also give an illuminating interpretation of the KSW estimator for oscillations in position space.

\section{Oscillations in the primordial bispectrum}

\subsection{Bispectrum shape and experimental constraints}

The general translation and rotation invariant primordial bispectrum of the curvature potential $\Phi$ can be written as
\begin{equation}
\label{eq_primbispectrum1} 
\langle \Phi(\mathbf{k_1}) \Phi(\mathbf{k_2}) \Phi(\mathbf{k_3}) \rangle = (2\pi)^3 \delta(\mathbf{k_{1,2,3}}) f_{NL} B_\Phi(k_1,k_2,k_3),
\end{equation}
where the bispectrum $B_\Phi$ is a function of the magnitude of the wave numbers $k_i$ and $f_{NL}$ is the amplitude of the bispectrum. The feature model bispectrum that we are primarily interested in is
\begin{equation}
\label{eq_oscispectrum1} 
B^{\mathrm{feat}}_\Phi(k_1,k_2,k_3) = \frac{6 \Delta_\Phi^2 f_{NL}}{(k_1k_2k_3)^2} \sin\left({\frac{2\pi K}{3 k_c} + \phi}\right).
\end{equation}
It oscillates linearly with the mean, $K= \frac{1}{3} (k_1+k_2+k_3)$, of the wave numbers. Here $\Delta_\Phi$ is the primordial power spectrum amplitude and the $1/k^6$ factor compensates for the phase space factor.  The bispectrum is parametrised by the amplitude $f_{NL}$ by the oscillation scale $k_c$ and by the phase $\phi$. The oscillation scale $k_c$ implies an efficient multipole periodicity of $l_c \simeq k_c \left[\tau - \tau_{rec}\right]$, where $\tau - \tau_{rec}$ is the conformal distance to recombination. 

Planck has searched for the feature bispectrum shape for sample frequencies in the range $0.01< k_c < 0.1$ at four different phases $\phi = 0,\pi/4,\pi/2,3\pi/4$~\cite{Ade:2013ydc}. Here, $k_c=0.01$ corresponds to an effective multipole periodicity $l_c = 140$. The best fit model has $k_c = 0.0185$ ($l_c = 260$) and phase $\Phi = 0$ with a significance of $3 \sigma$. This may correspond to weak hints for oscillation in the full bispectrum reconstruction, which were found for $l<500$. However, the statistical significance becomes much lower when one takes the number of statistically uncorrelated feature models into account that were searched. We note that the range of the oscillation frequency was constrained because of the limitations of the analysis method. With the present work, we target the unexplored range $k_c<0.01$ in particular.

\subsection{Separability}

For convenience, we rewrite the feature model (\ref{eq_oscispectrum1}) as a sum of sine and cosine contributions as
\begin{equation}
\label{eq_oscispectrum3} 
B^{\mathrm{feat}}_\Phi(k_1,k_2,k_3) = \frac{6 \Delta_\Phi^2}{(k_1k_2k_3)^2} \left[ f_1 \sin\left(\frac{2\pi K}{3 k_c}\right)  + f_2 \cos\left(\frac{2\pi K}{3 k_c}\right) \right],
\end{equation}
where $f_{NL} = \sqrt{f_1^2 + f_2^2}$ and $\Phi = \arctan{(\frac{f_2}{f_1})}$. This can be written in separated form as
\begin{align}
\label{eq_oscispectrum4} 
B^{\mathrm{feat}}_\Phi(k_1,k_2,k_3) = ~&6 f_1 \big[ -X(k_1) X(k_2) X(k_3) + \big(X(k_1) Y(k_2) Y(k_3) \nonumber \\
  &+ \textrm{2 perm.}\big) \big] + 6 f_2 \big[  Y(k_1) Y(k_2) Y(k_3) \nonumber \\
&- \big( X(k_1) X(k_2) Y(k_3) + \textrm{2 perm.}\big) \big],
\end{align}
where we have defined 
\begin{equation}
\label{eq_factorfunctions1} 
X(k) = \frac{\Delta_\Phi^{2/3}}{k^2} \sin\left(\frac{2\pi k}{3 k_c}\right) \hspace{0.5cm} \mathrm{and} \hspace{0.5cm} Y(k) = \frac{\Delta_\Phi^{2/3}}{k^2} \cos\left(\frac{2\pi k}{3k_c}\right).
\end{equation}
This separability property allows efficient computation and estimation of the bispectrum. One may also include an exponential decay factor of the form $exp(-\frac{K}{\mu})$, while retaining separability with identical formulas up to trivial replacements.

\section{Oscillations in the CMB bispectrum}

\begin{figure}
\resizebox{\hsize}{!}{
\includegraphics{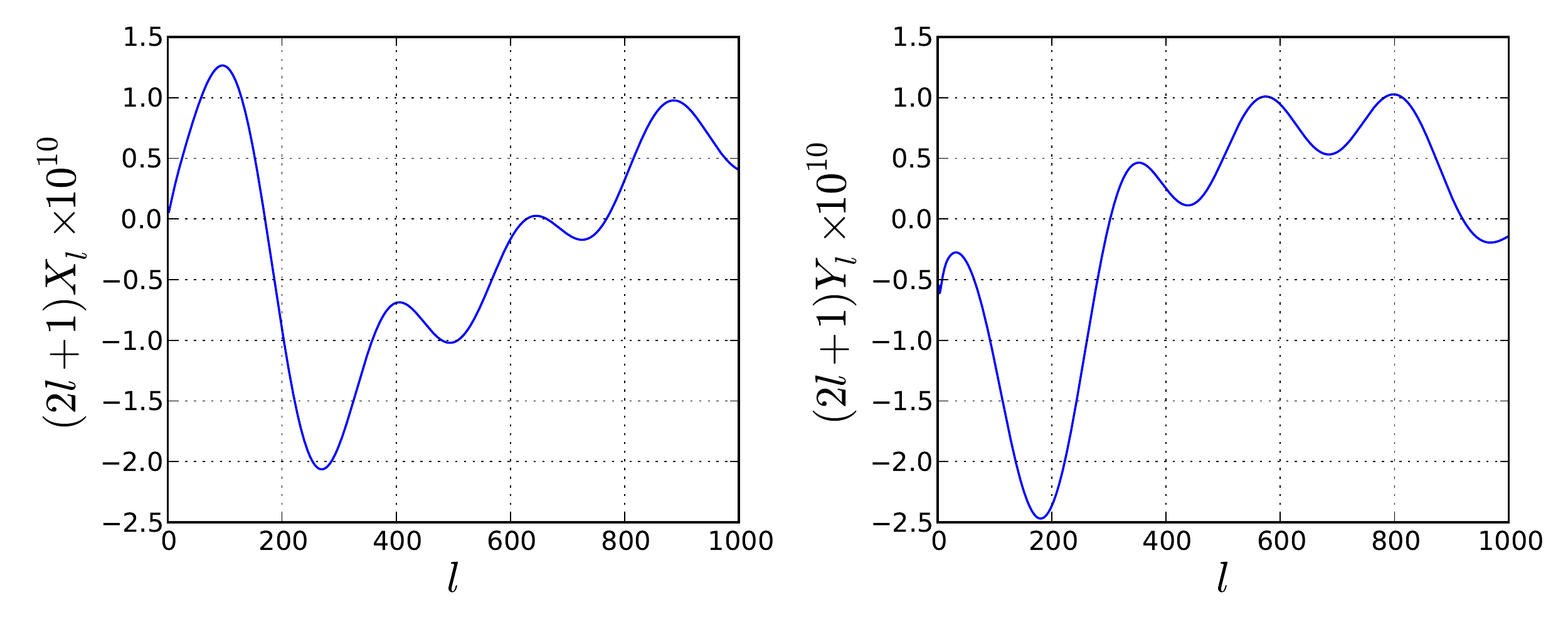}
}
\resizebox{\hsize}{!}{
\includegraphics{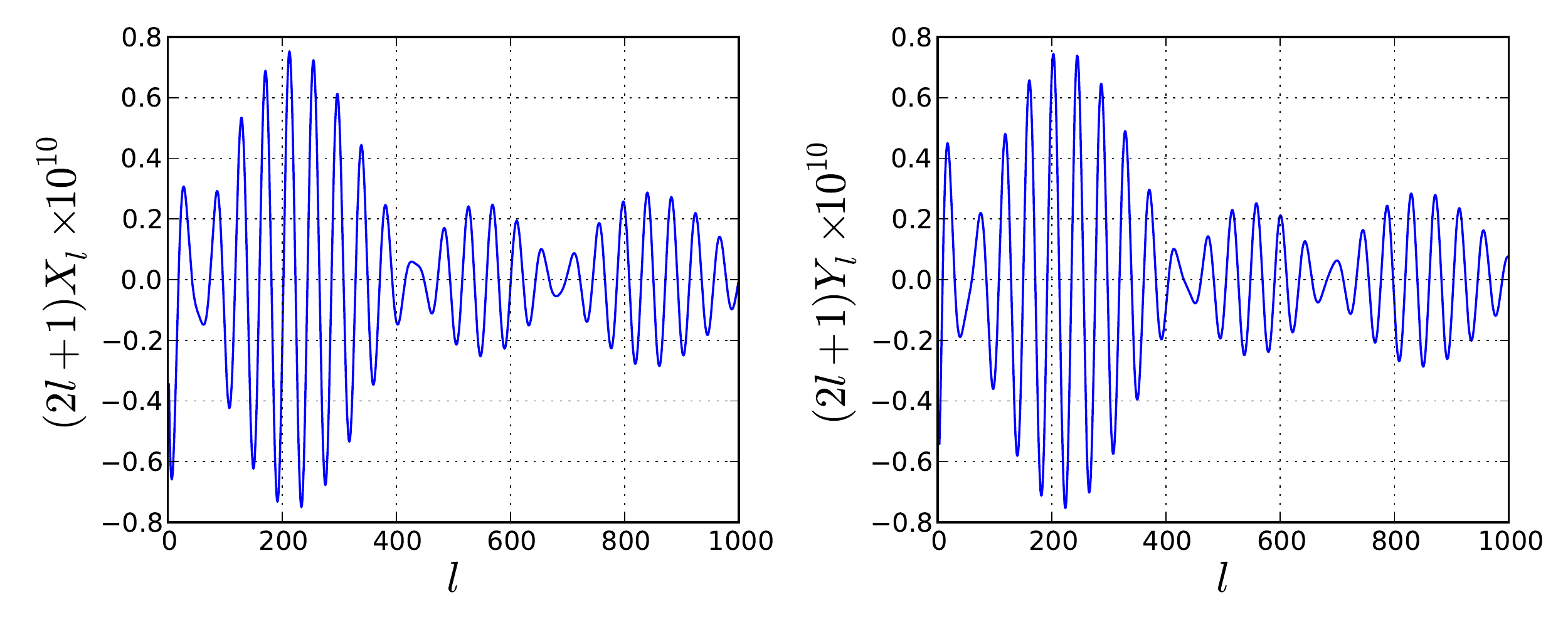}
}
\caption{Function $X$ and $Y$ of the feature model for
  $k_c = 0.01$ (top) and $k_c = 0.001$ (bottom) as a function of $l$ for $r=\tau_0-\tau_{rec}$. $X$ and $Y$ have units of Mpc$^{-1}$.} 
\label{fig:factorfunctions}
\end{figure}

From the separable expression (\ref{eq_oscispectrum4}) for the primordial bispectrum, one can calculate the CMB bispectrum with the standard line-of-sight integration method~\cite{Seljak:1996is}. The reduced bispectrum is then
\begin{align}
\label{eq_reducedbispectrum1} 
b_{l_1l_2l_3} = ~&6 f_1 \int dr~r^2~ \big[-X_{l_1}(r) X_{l_2}(r) X_{l_3}(r) + ( X_{l_1}(r) Y_{l_2}(r) Y_{l_3}(r)  \nonumber\\
&+ \textrm{2 perm.}) \big]  + 6 f_2 \int dr~r^2~ \big[ Y_{l_1}(r) Y_{l_2}(r) Y_{l_3}(r)  \nonumber\\
&- (X_{l_1}(r) X_{l_2}(r) Y_{l_3}(r) + \textrm{2 perm.}) \big].
\end{align}
where we have defined the functions 
\begin{align}
\label{eq_reducedbispectrum2} 
X_l(r) &= \frac{2}{\pi} \int dk k^2 X(k) j_l(kr) \Delta_l(k)\nonumber\\
         &= \frac{2}{\pi} \int dk \Delta_\Phi^{2/3} \sin\left(\frac{2\pi k}{3k_c}\right)  j_l(kr) \Delta_l(k),\\
Y_l(r) &= \frac{2}{\pi} \int dk k^2 Y(k) j_l(kr) \Delta_l(k)\nonumber\\
         &= \frac{2}{\pi} \int dk \Delta_\Phi^{2/3} \cos\left(\frac{2\pi k}{3k_c}\right)  j_l(kr) \Delta_l(k),
\end{align}
and where $j_l$ are spherical Bessel functions and $\Delta_l$ are the CMB transfer functions that we evaluate numerically with CAMB~\cite{Lewis:1999bs}. We note that these functions depend on $k_c$ and have to be evaluated for each oscillation frequency of interest. Both the transfer functions $\Delta_l(k)$ (in k space) and the Bessel functions $j_l$ are highly oscillatory integrals, which must be evaluated with sufficient sampling. Examples of the $X(r)$ and $Y(r)$ functions are given in figure \ref{fig:factorfunctions}.

\begin{figure}
\resizebox{\hsize}{!}{
\includegraphics{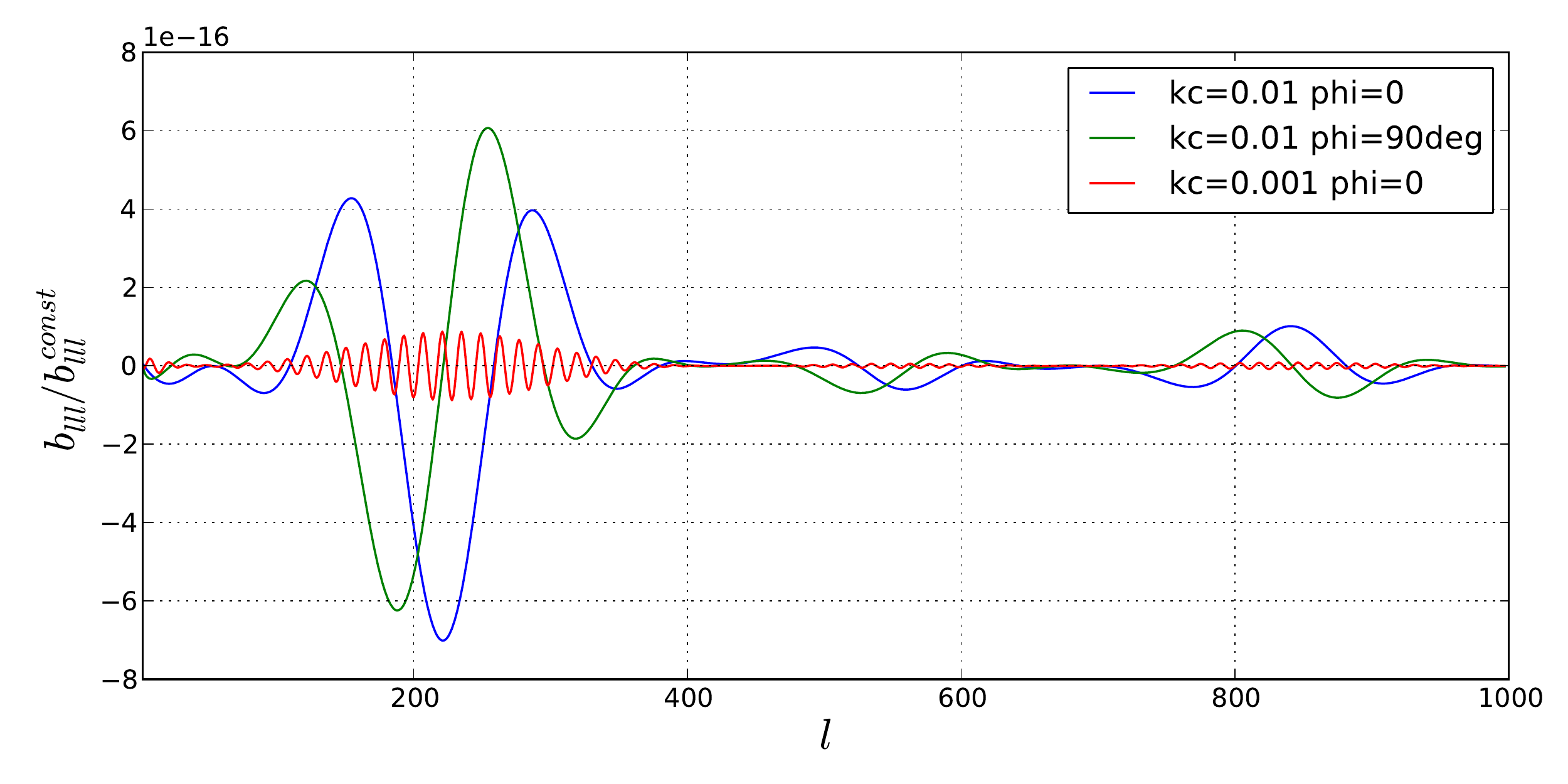}
}
\caption{Equal $l$ reduced bispectrum normalised by the large angle solution of the constant bispectrum for feature models with different parameters.} 
\label{fig:b_equall}
\end{figure}

To calculate the resulting CMB bispectrum, we perform the $r$ integral in equation (\ref{eq_reducedbispectrum1}). We choose a quadrature of about 2000 points in $r$ with a higher sampling in the range of recombination. The contribution of different values of $r$ is examined in more detail in section~\ref{sec:rintegral}. To plot the bispectrum, it is convenient to normalise by the constant bispectrum with natural $k^{-6}$ scaling, as proposed in~\cite{Fergusson:2008ra}. The constant primordial bispectrum is given by
\begin{equation}
\label{eq_blconst} 
B_{\Phi}^{const}(k_1,k_2,k_3) = \frac{1}{(k_1k_2k_3)^2}, 
\end{equation}
and its large angle Sach-Wolfe CMB solution is~\cite{Fergusson:2008ra}
\begin{align}
\label{eq_blconstsw} 
b^{const}_{l_1l_2l_3} = \left(\frac{1}{3}\right)^3 \frac{1}{(2l_1+1) (2l_2+1) (2l_3+1)} \nonumber\\
\times \left[ \frac{1}{l_1 +l_2+l_3+3} + \frac{1}{l_1+l_2+l_3} \right].
\end{align}
In figure \ref{fig:b_equall}, we show a simple 1-dimensional visualisation, the equal $l$ bispectrum $b_{lll}$ normalised by $b^{const}_{lll}$ for different frequencies and phases of the feature model. 

\section{The KSW estimator for the oscillating bispectrum}
\label{sec:ksw}

\subsection{KSW estimator}

The optimal KSW estimator for a sum of bispectra $b^i$ in the presence of a mask and non-uniform noise is~(\cite{Komatsu:2003iq,Babich:2005en,Komatsu:2010hc})
\begin{equation}
\label{eq_kswphase1} 
f_i= \sum_j (F^{-1})_{ij} S_j,
\end{equation}
with 
\begin{align}
\label{eq_kswphase2} 
S_i =~&\frac{1}{6} \sum_{lm} G^{m_1m_2m_3}_{l_1l_2l_3}   b^i_{l_1l_2l_3} \Big[ (C^{-1}a)_{l_1m_1} (C^{-1}a)_{l_2m_2} (C^{-1}a)_{l_3m_3} \nonumber\\
&- 3 (C^{-1})_{l_1m_1,l_2m_2} (C^{-1}a)_{l_3m_3} \Big]
\end{align}
and with the Fisher matrix given by
\begin{align}
\label{eq_kswphase3} 
F_{ij} =~& \frac{1}{6} \sum_{lm} \sum_{l'm'}  G^{m_1m_2m_3}_{l_1l_2l_3}  b^i_{l_1l_2l_3} (C^{-1})_{l_1m_2,l'_1m'_1} (C^{-1})_{l_2m_2,l'_2m'_2} \nonumber\\
&\times(C^{-1})_{l_3m_3,l'_3m'_3}  b^j_{l'_1l'_2l'_3} G^{m'_1m'_2m'_3}_{l'_1l'_2l'_3}.
\end{align}

In the present case of an oscillation with a single scale $k_c$, the bispectrum is a sum $b = f_1 b^1 + f_2 b^2$ of sine and cosine components, as given by equation (\ref{eq_reducedbispectrum1}).

To numerically evaluate the KSW estimator terms $S_i$ we need the weighted maps
\begin{align}
\label{eq_ksw1} 
M_X(r, \hat{n}) =& \sum_{lm} (C^{-1}a)_{lm}X_l(r) Y_{lm}(\hat{n}),\nonumber\\
M_Y(r, \hat{n}) =& \sum_{lm} (C^{-1}a)_{lm}Y_l(r) Y_{lm}(\hat{n}).
\end{align}
The cubic KSW estimator, which is exact in the case of a full sky observation, is given in terms of these maps by
\begin{align}
\label{eq_ksw2} 
S^{cub}_1 &=  \int r^2 dr \int d\Omega  \left[ -M^3_X(r,\hat{n}) +\left(3 M_X(r,\hat{n}) M_Y(r,\hat{n}) M_Y(r,\hat{n})\right) \right], \nonumber\\
S^{cub}_2 &=  \int r^2 dr \int d\Omega  \left[ M^3_Y(r,\hat{n}) -\left(3 M_X(r,\hat{n}) M_X(r,\hat{n}) M_Y(r,\hat{n})\right) \right].
\end{align}

Partial sky coverage can be taken into account by incorporating the linear term of the KSW estimator, so that $S_i = S^{cub}_i + S^{lin}_i$ with
\begin{align}
\label{eq_ksw3} 
S_1^{lin} =  &-3\int r^2 dr \int  d\Omega  \bigl[- M_X(r,\hat{n}) \left< M^2_X(r,\hat{n}) \right>  \nonumber\\
&+ M_X(r,\hat{n}) \left< M^2_Y(r,\hat{n}) \right> + 2 M_Y(r,\hat{n}) \left< M_X(r,\hat{n}) M_Y(r,\hat{n}) \right> \bigr], \nonumber\\
S_2^{lin} =  &-3\int r^2 dr \int d\Omega  \bigl[ M_Y(r,\hat{n}) \left< M^2_Y(r,\hat{n}) \right> \nonumber\\ 
&- M_Y(r,\hat{n}) \left< M^2_X(r,\hat{n}) \right> - 2 M_X(r,\hat{n}) \left< M_X(r,\hat{n}) M_Y(r,\hat{n}) \right>\bigr],
\end{align}
where the expectation values have to be evaluated by Monte Carlo averaging over Gaussian realisations drawn with the same beam, mask, and noise properties as expected in the data. To make the KSW estimator optimal for a non-uniform sky coverage, it is necessary to perform an inverse covariance weighting with the non-diagonal covariance matrix. This is a computationally challenging problem~(\cite{Smith:2009jr,Elsner:2012dj}). It was noted in~\cite{Ade:2013ydc} that one can also achieve excellent results by assuming a diagonal covariance matrix $\hat{C}_l=C_l+N_l$, where $N_l$ assumes homogeneous noise, and by using a diffusive inpainting on the masked areas. In this approximation, the Fisher matrix scales proportionally to the visible fraction of the sky $f_{sky}$.

For the remainder of this paper, we assume full sky coverage so that
\begin{align}
\label{eq_kswphase8} 
S_i &= \frac{1}{6} \sum_{lm} \frac{G^{m_1m_2m_3}_{l_1l_2l_3}   b^i_{l_1l_2l_3}}{C_{l_1}C_{l_2}C_{l_3}} a_{l_1m_1}a_{l_2m_2}a_{l_3m_3}, \nonumber\\
F_{ij} &= \frac{1}{6} \sum_{l} I_{l_1l_2l_3}  \frac{  b^i_{l_1l_2l_3} b^j_{l_1l_2l_3} }{C_{l_1}C_{l_2}C_{l_3}}, 
\end{align}
where $I_{l_1l_2l_3} = \sum_{\mathrm{all~m}} (G^{m_1m_2m_3}_{l_1l_2l_3})^2$.

From the inverse Fisher matrix (i.e. the covariance), one obtains the correlation between the sine and cosine terms. For example, for $l=1000$ and $k_c=0.01$, the correlation matrix is 
\begin{equation}
\label{eq_kswphase4} 
corr(b^i,b^j) =  \frac{F^{-1}_{ij}}{\sqrt{F^{-1}_{ii}F^{-1}_{jj}}} = 
\begin{pmatrix}
 1 & -0.04\\
 -0.04 & 1
\end{pmatrix}
\end{equation}
which shows a weak correlation as expected.  From $f_1$ and $f_2$ of eq. (\ref{eq_reducedbispectrum1}), we can calculate the amplitude and phase of the oscillation. The variance of the quantities $f_{NL}$ and $\Phi$ can be calculated by error propagation from $f_1$ and $f_2$. For example
\begin{equation}
\label{eq_kswphase5} 
\sigma_f^2 = \frac{f_1^2}{f^2} \sigma_{f_1}^2 + \frac{f_2^2}{f^2} \sigma_{f_2}^2 + 2 \frac{f_1 f_2}{f^2} cov(f_1,f_2). 
\end{equation}
In the approximation of a diagonal covariance matrix and $\sigma_{f_1} = \sigma_{f_2}$, this gives $\sigma_f = \sigma_{f_1}$. In this approximation the sensitivity on the phase depends on $f$ as $\sigma_\Phi = \frac{\sigma_{f_1}}{\sqrt{f}}$.

\begin{figure}
\resizebox{\hsize}{!}{
\includegraphics{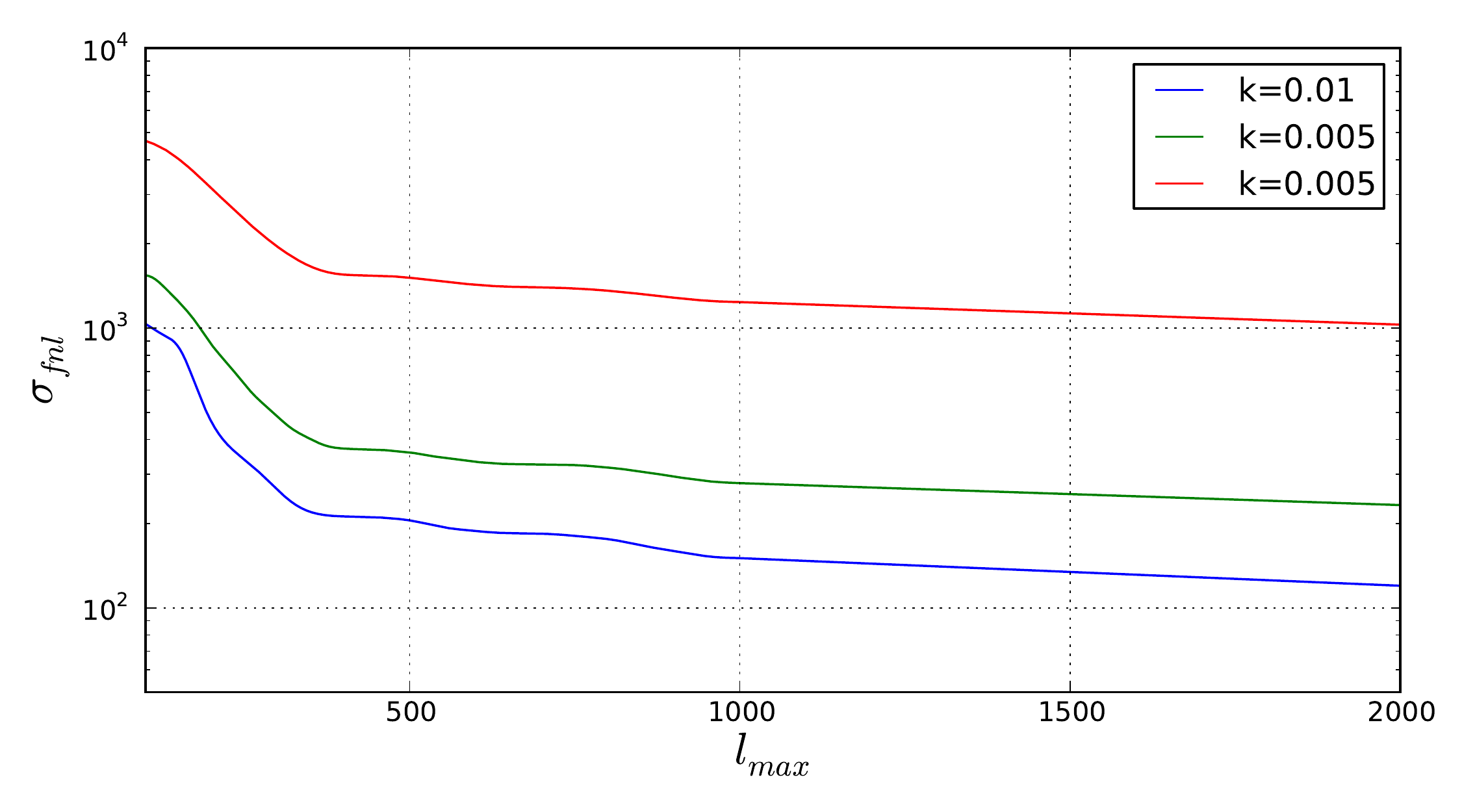}
}
\caption{Fisher forecast of $\sigma_{f_{NL}}$ for the feature model for different $k_c$, assuming a noiseless full sky experiment.} 
\label{fig:fisherforecast}
\end{figure}

The Fisher matrix allows us to forecast the precision that can be obtained on the bispectrum parameters $f_1,f_2$. With the approximation that the $f_1,f_2$ covariance matrix is a multiple of the unit matrix, the Fisher forecast on $f_1,f_2$ equals the forecast on $f_{NL}$. The precision on $f_{NL}$ is then given by $\sigma_{f_{NL}} = \frac{1}{\sqrt{F}}$. For a noiseless full-sky experiment, the Fisher forecast for the feature model is shown in figure \ref{fig:fisherforecast} for a number of different oscillation frequencies.

\subsection{Estimating the frequency}

\begin{figure}
\centering
\resizebox{0.8\hsize}{!}{
\includegraphics{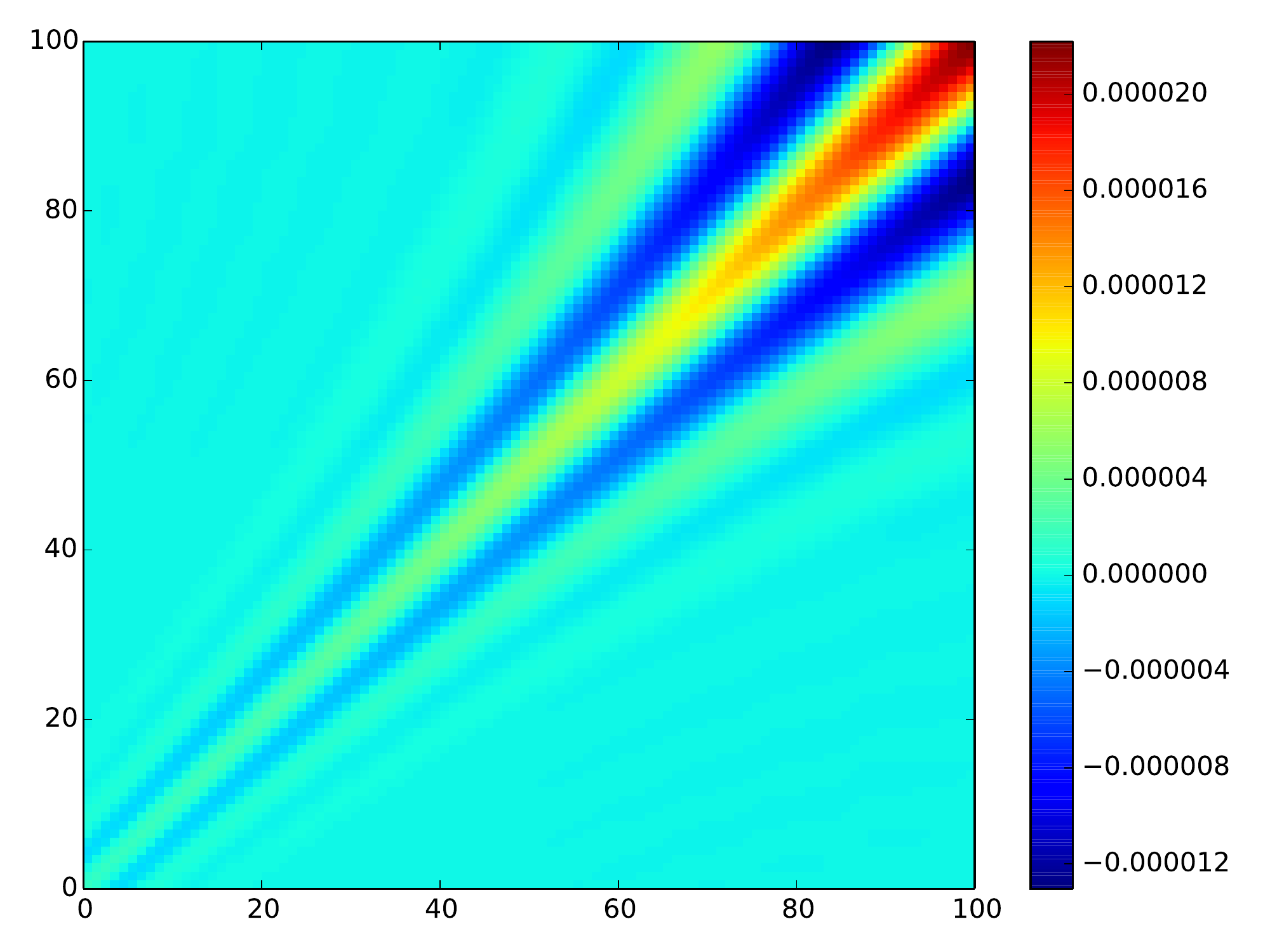}
}
\resizebox{0.8\hsize}{!}{
\includegraphics{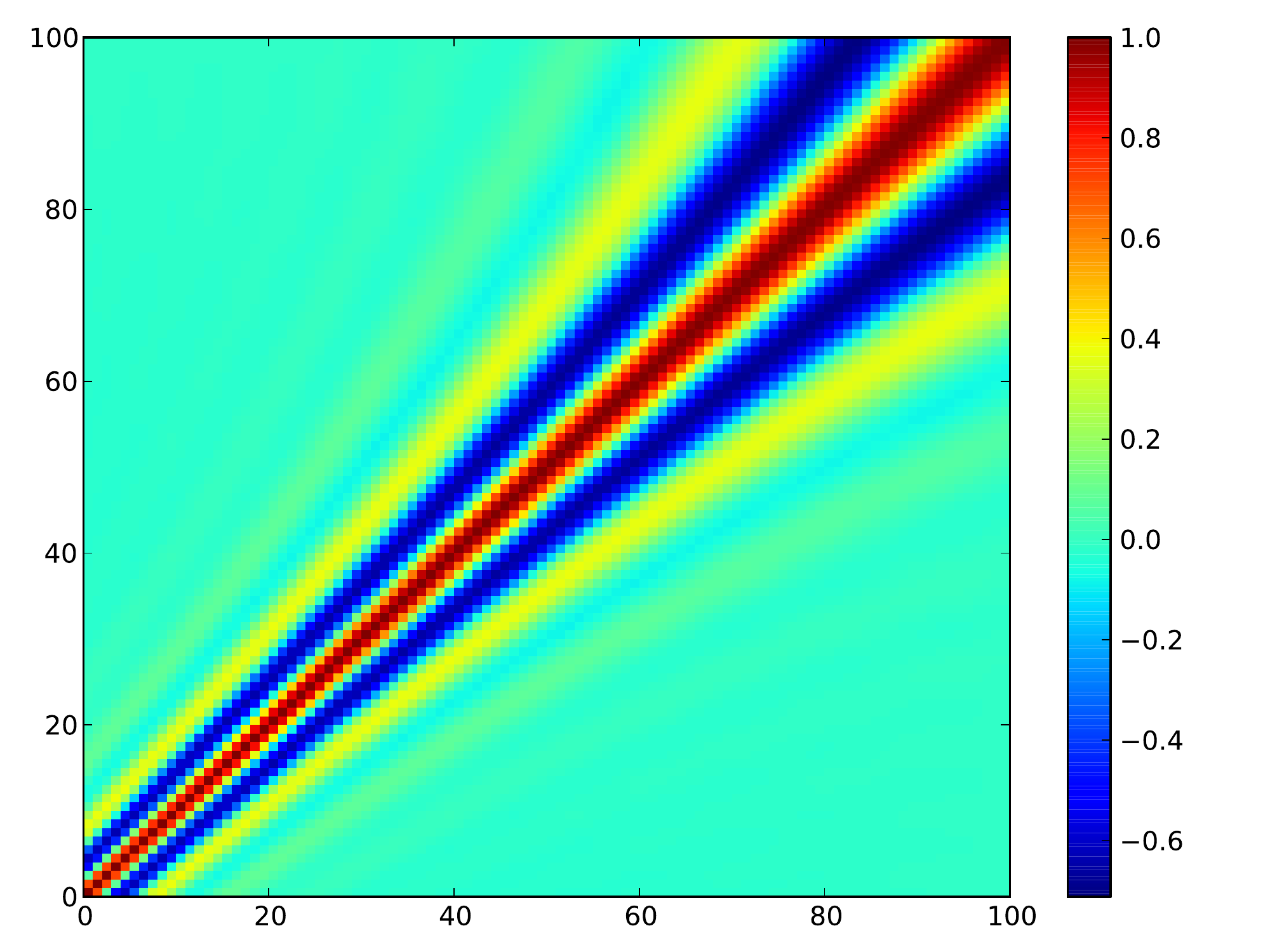}
}
\resizebox{0.8\hsize}{!}{
\includegraphics{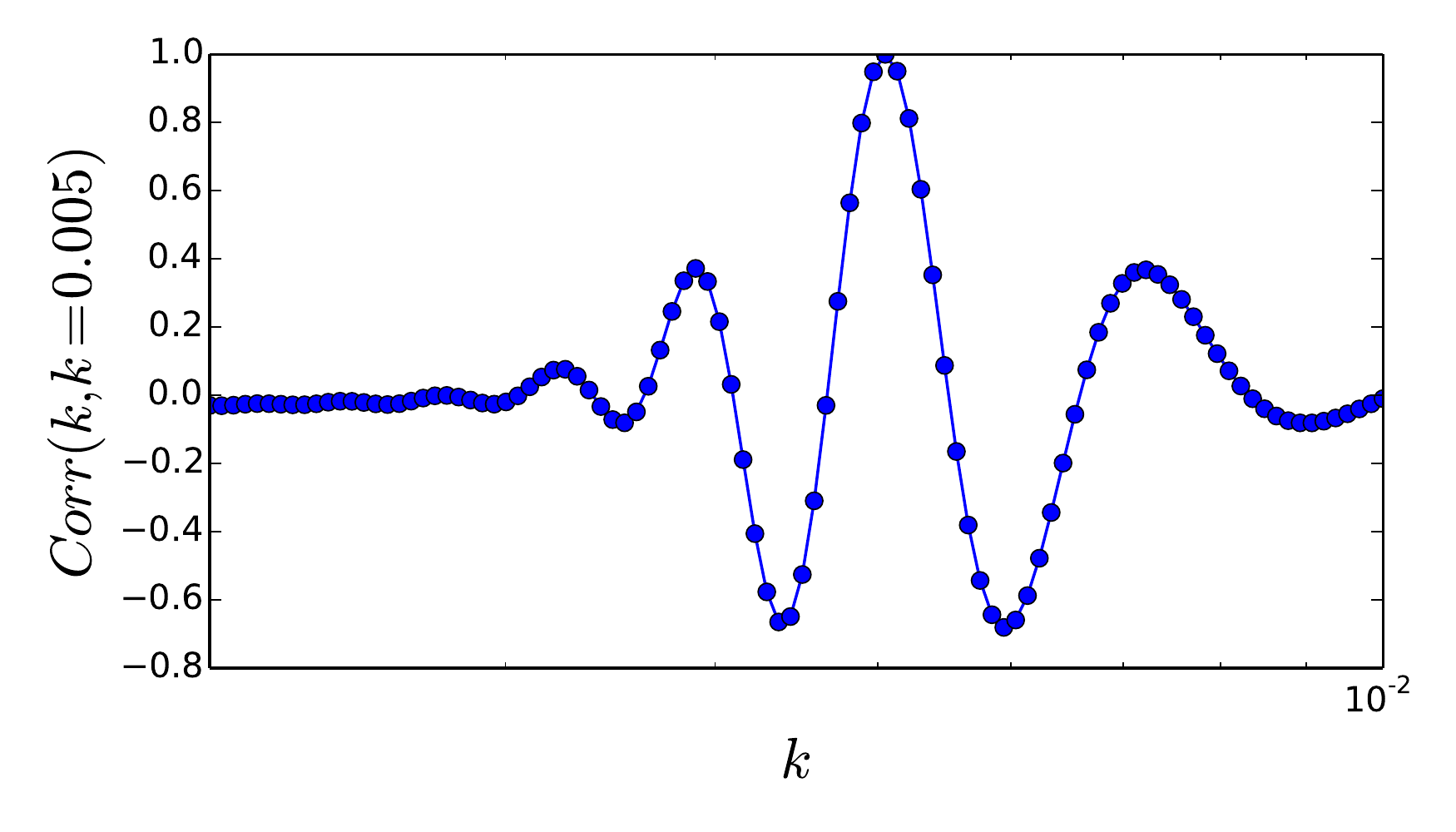}
}
\caption{Top: Fisher matrix $F_{ij}$ for 100 logarithmically spaced frequencies between $k_c=0.002$ and $k_c=0.01$. Middle: Corresponding correlation matrix $corr(f'_i,f'_j)$. Bottom: Correlation matrix one-dimensional slice $corr(f_i,f_j)$ for $k_i=0.005$.} 
\label{fig:fishermatrix}
\end{figure}

The estimator described above explicitly estimates the amplitude $f_{NL}$ and phase $\phi$ of the oscillation for a fixed frequency $k_c$. To estimate the oscillation frequency $k_c$, it is necessary to sample the frequency space with the KSW estimator and search for peaks in the significance of the estimated amplitude. We assume the primordial bispectrum to be given by a single oscillation frequency and not a spectrum of contributions. We consider only the sine component of the bispectrum first, meaning that the phase is $\phi=0$ (see below for the generalisation to the phase). In this case, the estimator for a frequency $k_i$ is
\begin{equation}
\label{eq_kc3} 
\hat{f}_i= (F^{-1})_{ii} S_i
\end{equation}
with covariance (in the usual Gaussian approximation)
\begin{align}
\label{eq_kc4} 
cov(\hat{f}_i,\hat{f}_j) =& \big<\hat{f}_i \hat{f}_j\big> - \big<\hat{f}_i\big> \big<\hat{f}_j\big>\\
 =& \frac{F_{ij}}{F_{ii}F_{jj}}.
\end{align}
The Fisher matrix is given by equation (\ref{eq_kswphase3}), where the index $i$ now goes over frequency sampling points. An example of the Fisher matrix $F_{ij}$ is shown in figure \ref{fig:fishermatrix} (top). The corresponding correlation matrix is $corr(\hat{f}_i,\hat{f}_j) = \frac{F_{ij}}{\sqrt{F_{ii} F_{jj}} }$, shown in figure \ref{fig:fishermatrix} (middle). 

A one-dimensional slice of the correlation matrix is shown in figure \ref{fig:fishermatrix} (bottom) for $k_c=0.005$. The plot shows strong anti-peaks to both sides of the maximum and several small secondary peaks. This is not an artefact of the chosen estimator, but the physical overlap of the CMB bispectra induced by different primordial oscillation frequencies. The frequency sampling must be at least sufficient to resolve the peak structure of this plot. However, the width of the central peak does not directly limit the precision $\sigma_k$ with which the primordial frequency can be determined. 

To evaluate the precision with which $k_c$ can be estimated, we note that the correlation matrix is identical to the bispectrum correlator,
\begin{equation}
\label{eq_corr} 
C(B,B')=\frac{1}{\mathcal{N}(B) \mathcal{N}(B')} \sum_{l} \frac{  B^i_{l_1l_2l_3} B^j_{l_1l_2l_3} }{C_{l_1}C_{l_2}C_{l_3}}, 
\end{equation}
which is the usual measure to discriminate bispectra (see e.g.~\cite{Fergusson:2008ra}). It gives the estimated proportion of $f$ that is recovered when estimating a spectrum $B'$ when the true underlying spectrum is $B$. If the underlying bispectum is $f_{k_c} B_{k_c}$, the expectation value of the estimated amplitude at a different frequency $k$ is thus $\left<\hat{f}(k)\right> = f_{k_c} Corr(B_{k_c},B_{k}$). The variance at each data point is independent of the signal. Thus, the correlation matrix approximates the estimates in the $k$ range where the signal dominates the variance (compare figure \ref{fig:fishermatrix} with the estimation example in figure \ref{fig:kcsweep}). 

Knowing the means $f(k)$ and the covariance matrix $Cov(k_1,k_2)$ and using the property that each estimator $\hat{f}(k)$ is Gaussian distributed, we can write the continuum likelihood for the estimated amplitudes $\hat{f}(k)$, if the true bispectrum is $f_{k_c} B_{k_c}$: 
\begin{align}
\label{eq_kclikeli} 
-2 &\ln ~\mathcal{L}(\hat{f}(k)|f_{k_c},k_c) = \ln|Cov| \nonumber \\
&+\left[ \int dk_1 dk_2 ~(\hat{f}(k_1) - \mu(k_1)) ~ Cov^{-1}(k_1,k_2) ~ (\hat{f}(k_2) - \mu(k_2)) \right],
\end{align}
with mean 
\begin{equation}
\label{eq_kcmean} 
\mu(k) = f_{k_c} Corr(B_{k_c},B_{k}).
\end{equation}
This likelihood could be explored by Monte Carlo to find maximum likelihood estimates of $f_{k_c}$ and $k_c$ if a significant peak of $f_{NL}$ is found in the spectrum. The Fisher matrix to forecast optimal precision on the estimated parameters is 
\begin{equation}
\label{eq_fishercont} 
F_{\theta \theta'} = \left< \frac{\partial \mathcal{L}}{\partial \theta \partial \theta'}\right> = \int dk_1 dk_2 \frac{\partial \mu(k_1)}{\partial \theta} Cov^{-1}(k_1,k_2) \frac{\partial \mu(k_2)}{\partial \theta'},
\end{equation}
where $\theta \in \{f_{k_c},k_c\}$. This can be integrated numerically for any given fiducial parameters.  

In the above discussion, we assumed that the bispectrum only has a sine contribution. The generalisation of eq. (\ref{eq_kclikeli}) to a free phase is
\begin{align}
\label{eq_kclikeli2} 
&-2 \ln ~\mathcal{L}(\hat{f}^{1,2}(k)|f_1,f_2,k_c) = \ln|Cov|  \nonumber \\
&+ \sum_{\substack{i=1,2\\ j=1,2}} \int dk_1 dk_2 ~(\hat{f}^i(k_1) - \mu^i(k_1)) ~ Cov_{ij}^{-1}(k_1,k_2) ~ (\hat{f}^j(k_2) - \mu^j(k_2)), 
\end{align}
with the mean
\begin{equation}
\label{eq_kcmean2} 
\mu^i(k) = f_{k_c} Corr(f_1 B^1_{k_c}+f_2 B^2_{k_c},B^i_{k}).
\end{equation}
The correlation matrix can be split into terms of $f_1$ and $f_2$ for efficient evaluation of the likelihood. 

As we have seen, the estimation of the frequency requires to run a large number of estimators (around 100 to cover the frequency interval $k_c=0.001$ to $k_c=0.01$). For each of these estimators, it is necessary to calculate the linear term in eq. (\ref{eq_ksw3}) via a Monte Carlo averaging procedure over Gaussian map realisations with the same mask and noise properties as present in the experiment. Convergence of the linear term is usually achieved with 100 Monte Carlo realisations, although several hundreds can be used to improve accuracy. After the optimisation of the conformal distance integral, that is reviewed in section \ref{sec:rintegral}, one can expect, with an angular resolution of $l_{max} = 2000$, a calculation time of about one day on a single cpu for a single frequency, including the Monte Carlo averaging. On a computation grid, one can thus easily cover the frequency range of interest.

\section{Map making}
\label{sec:mapmaking}

To verify the implementation and unbiased nature of the estimator, it is useful to be able to generate maps with the bispectrum signature of interest. The authors of~\cite{Smith:2006ud} introduced an algorithm that generates maps for arbitrary bispectra in the weak non-Gaussian limit. A map is constructed from a linear combination of a Gaussian and a non-Gaussian contribution as $a_{lm} = a^L_{lm} + f_{NL} a^{NL}_{lm}$. The straightforward implementation of the algorithm gives a non-Gaussian contribution of the form
\begin{align}
\label{eq_map1} 
a^{NL}_{lm} =~&  f_1 \int r^2 dr \Bigl[ - X_l(r) \int d\Omega ~Y_{lm}^* M^2_X(r,\hat{n}) \nonumber\\
&+\left( X_l(r) \int d\Omega ~Y_{lm}^* M_Y(r,\hat{n}) M_Y(r,\hat{n}) + 2 \mathrm{perm.} \right) \Bigr] \nonumber\\
&+ f_2 \int r^2 dr \Bigl[  Y_l(r) \int d\Omega ~Y_{lm}^* M^2_Y(r,\hat{n}) \nonumber\\
&-\left(Y_l(r) \int d\Omega ~Y_{lm}^* M_X(r,\hat{n}) M_X(r,\hat{n}) + 2 \mathrm{perm.} \right) \Bigr].
\end{align}
It is thus easy to create maps of arbitrary phase from the sine and cosine terms. 

\begin{figure}
\resizebox{\hsize}{!}{
\includegraphics{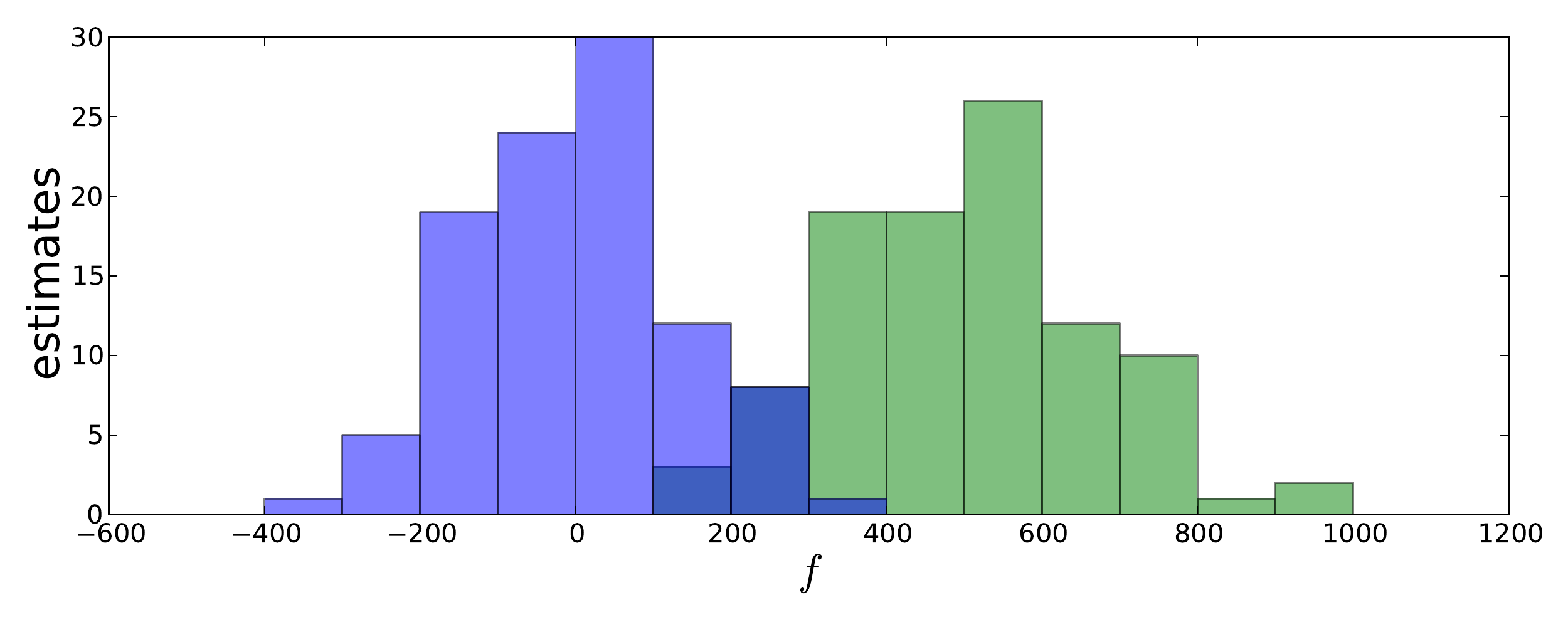}
}
\resizebox{\hsize}{!}{
\includegraphics{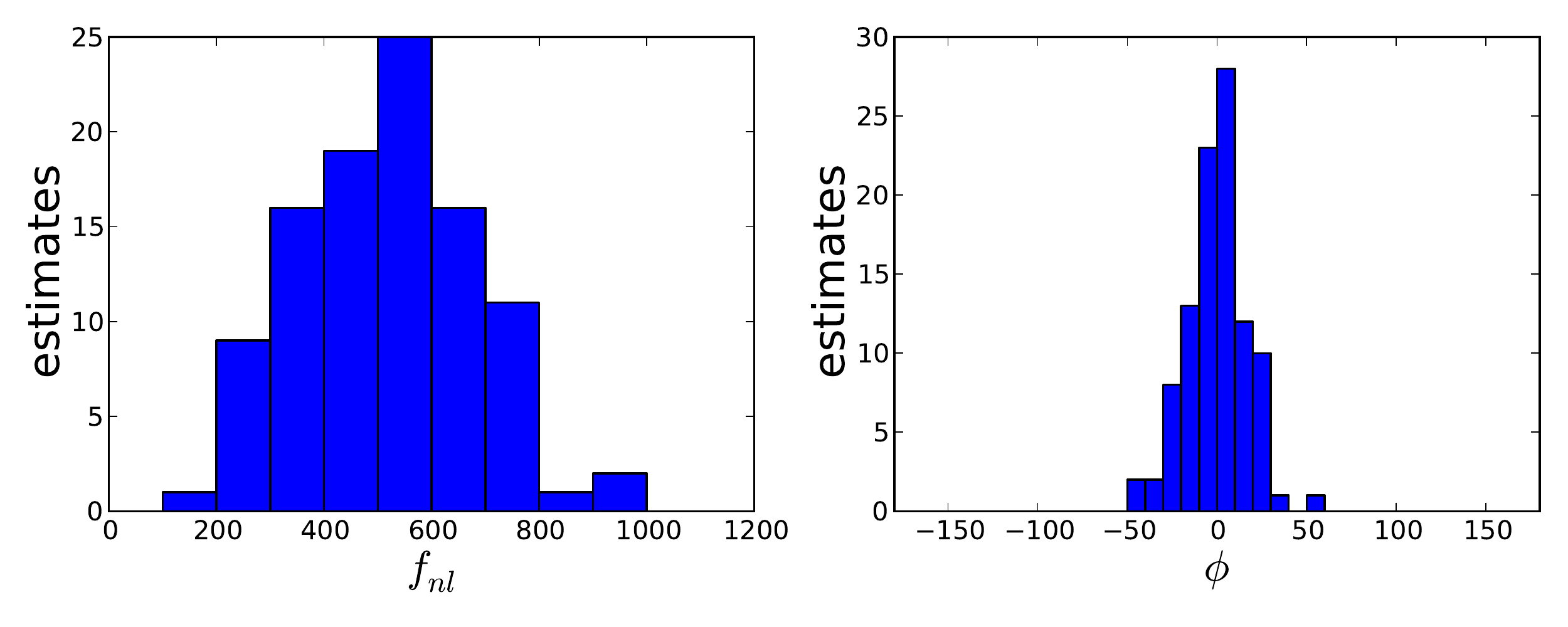}
}
\resizebox{\hsize}{!}{
\includegraphics{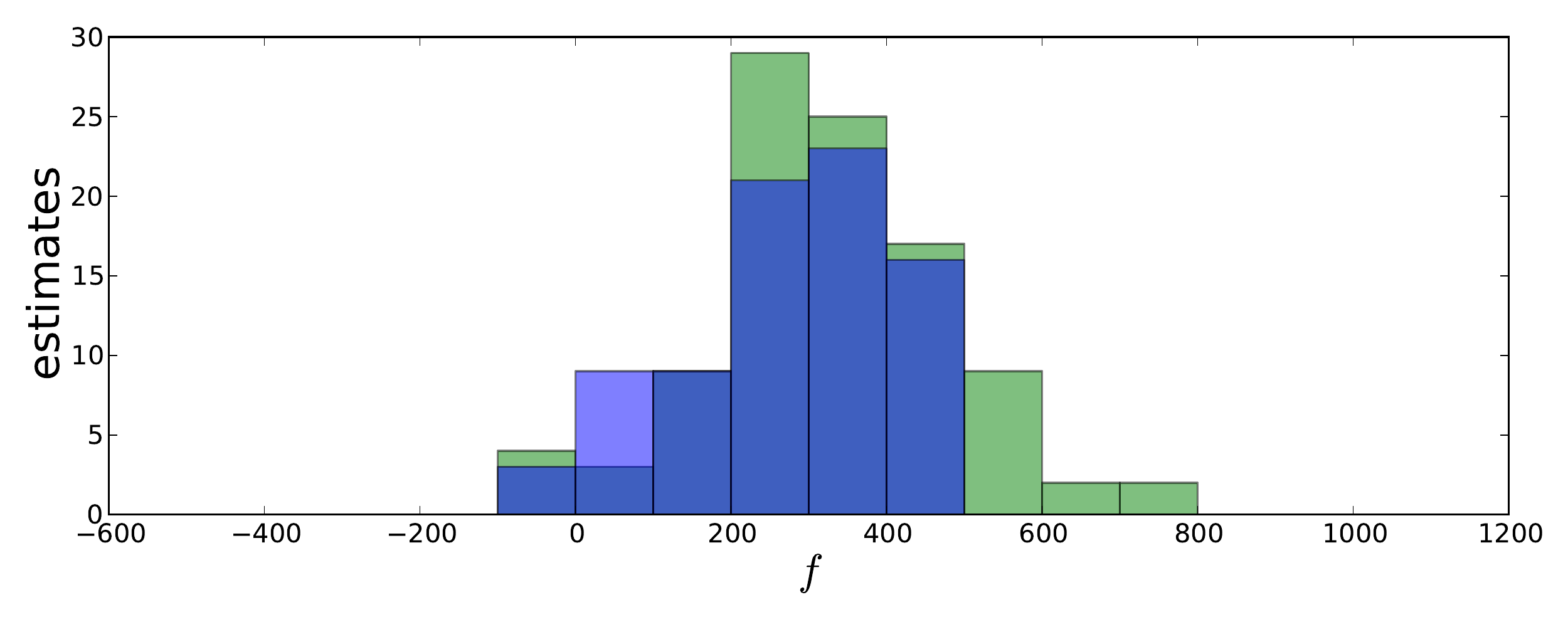}
}
\resizebox{\hsize}{!}{
\includegraphics{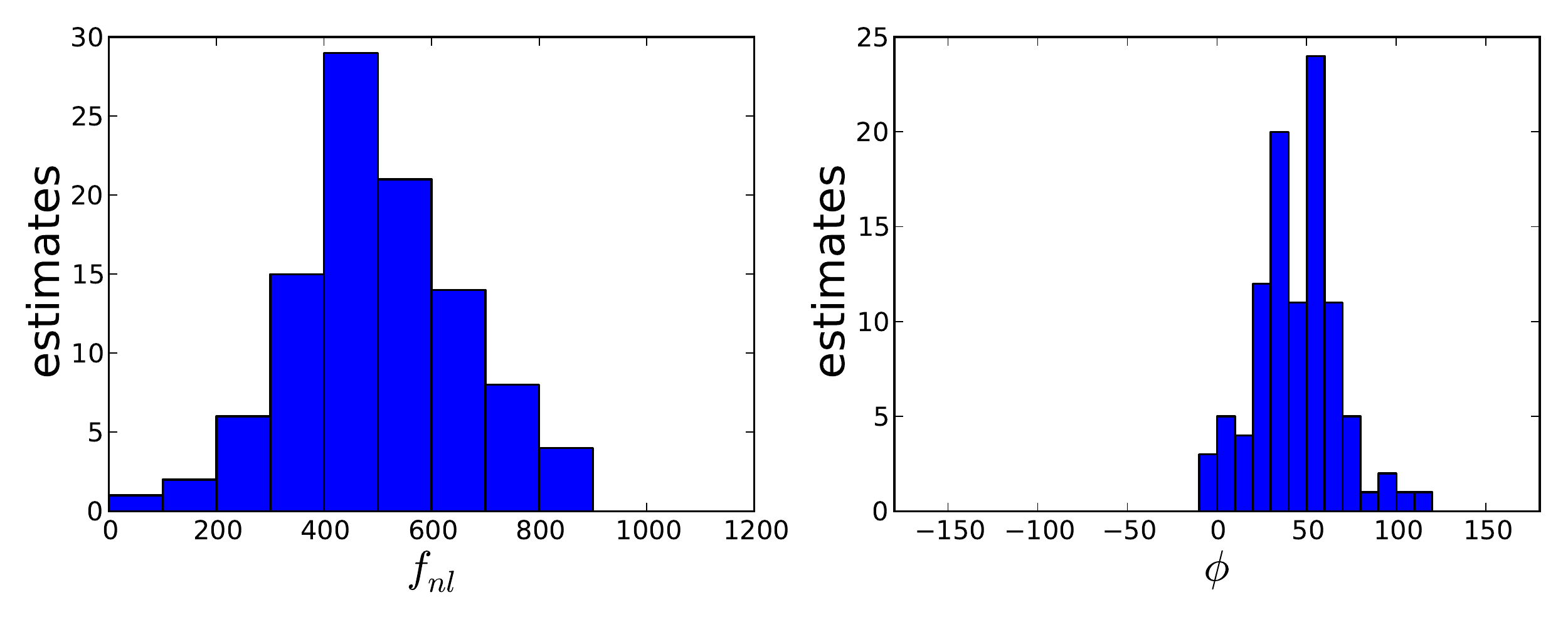}
}
\caption{Parameter estimation histograms for 100 maps with linear model non-Gaussianity with $k_c=0.01$ and $l_{max}=1000$. The maps were created with
  $f_{NL} = 500$ and $\phi=0^\circ$ (first two rows) and $\phi=45^\circ$ (rows three and four). Top: Reconstructed $f_1$ (green) and $f_2$ (blue) amplitudes for $\phi=0^\circ$. Second row: Corresponding reconstructed amplitudes $f_{NL}$ and phases $\phi$. Third and fourth row: Same as above but with phase $\phi=45^\circ$. } 
\label{fig:mapcreationestimation}
\end{figure}

An example of two sets of 100 maps, created and then estimated with the algorithms presented here, is shown in figure \ref{fig:mapcreationestimation}. The means and variances in these histograms are compatible with their Fisher forecast. We note that the distribution of $f_{NL}$ is not Gaussian but follows a Rayleigh distribution, since $f_{NL}$ represents the magnitude of the vector of the two directional components $f_1$ and $f_2$. 

\begin{figure}
\resizebox{\hsize}{!}{
\includegraphics{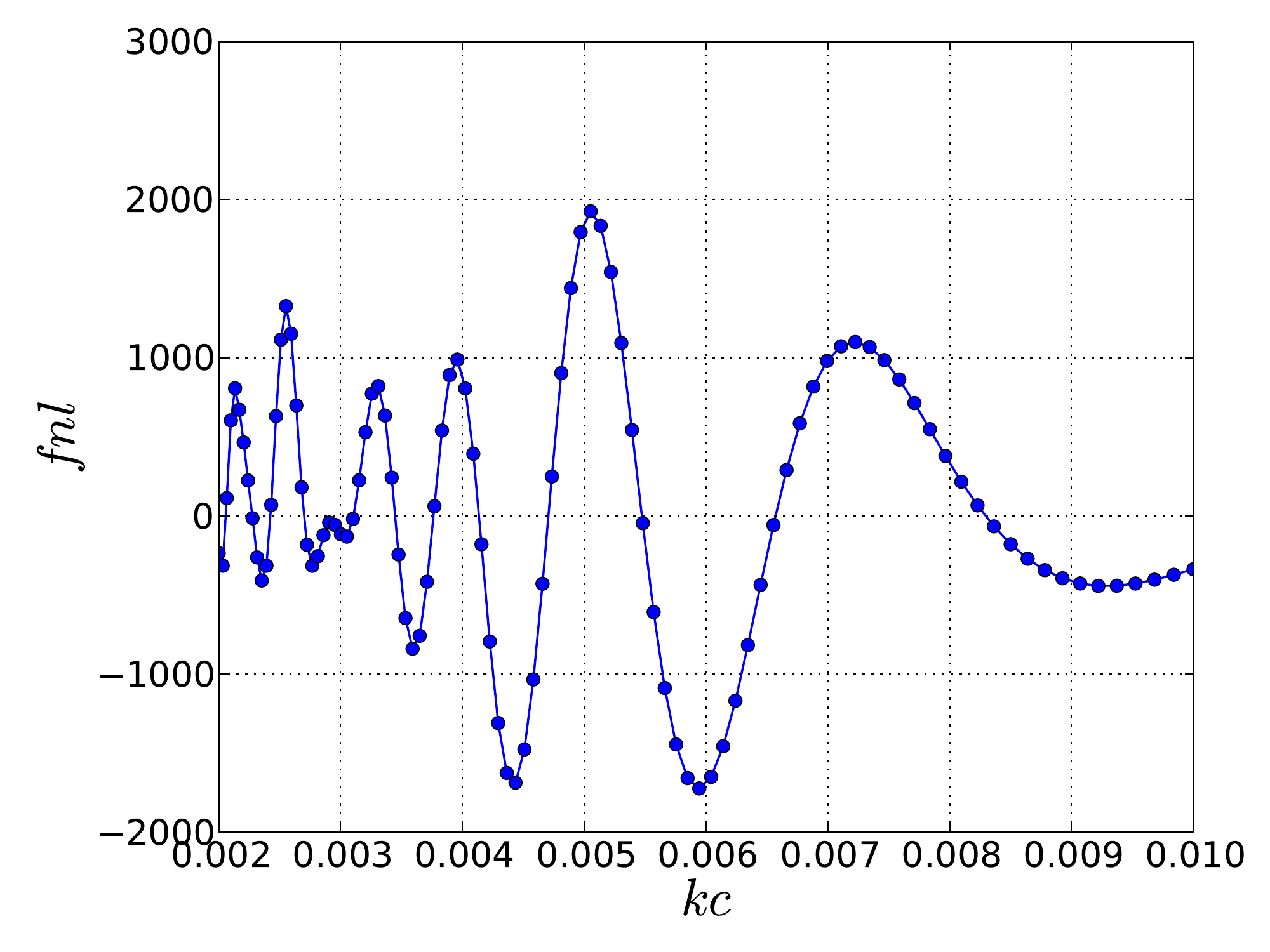}
\includegraphics{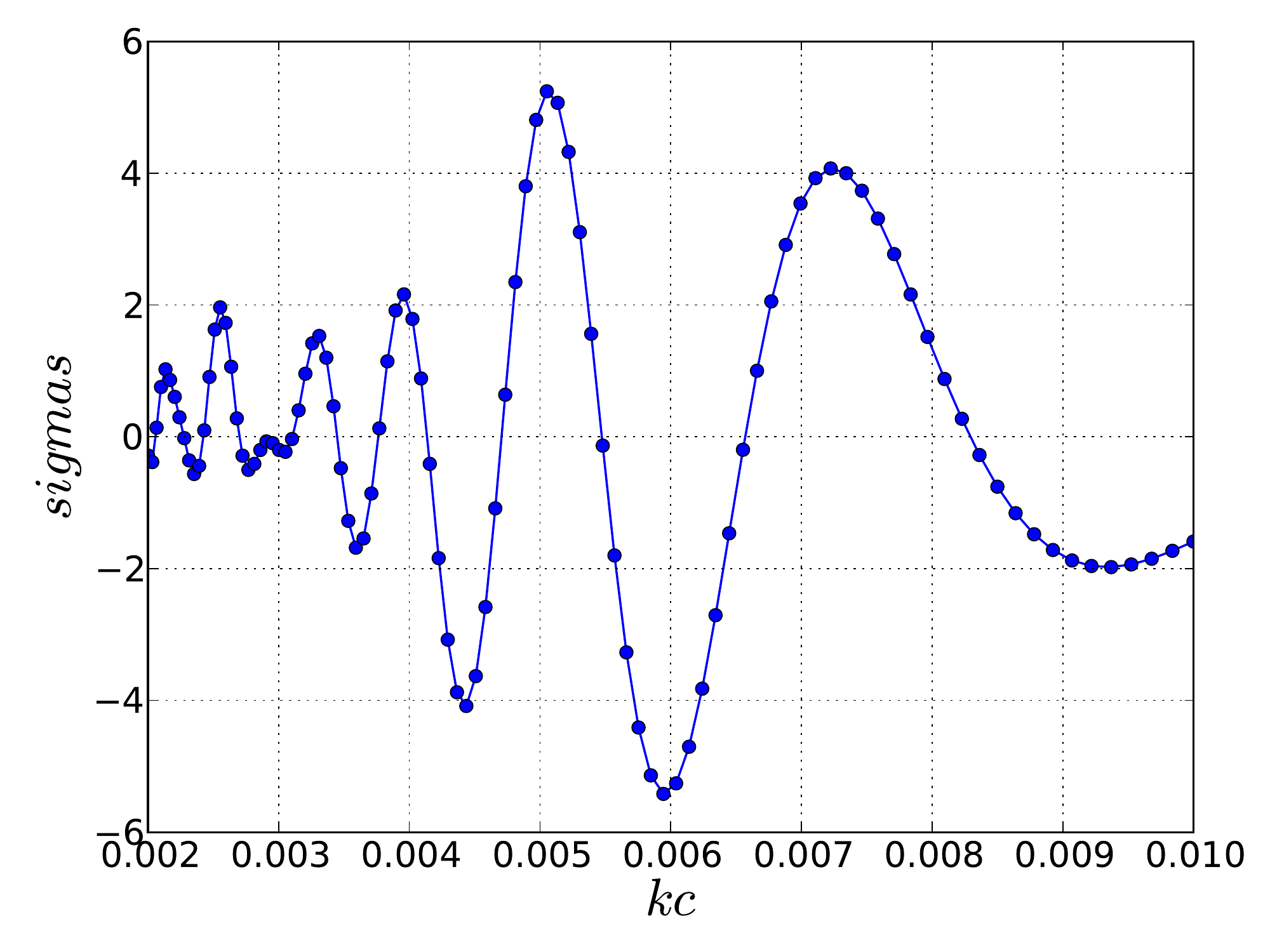}
}
\caption{Left: Frequency sweep over a map with $k_c = 0.005$ and $f_{NL}=2000$. Right: As left but in units of $\sigma_{f_{NL}}$.} 
\label{fig:kcsweep}
\end{figure}

An example of a frequency sweep can be found in figure \ref{fig:kcsweep} for the phase $\varphi = 0^\circ$. It shows the secondary peaks that are expected from the correlation of different frequencies.

\section{Computational speed improvement by optimisation of the r-integral}
\label{sec:rintegral}

Due to the large number of frequencies that have to be sampled with the corresponding estimator, the search for oscillations is computationally challenging. The time critical steps are the calculation of the Fisher matrix and the necessity of calculating a large number of $r$ dependent KSW filtered maps. The latter problem becomes even more severe if one has to estimate many Monte Carlo generated maps for the calculation of the linear term. The situation can be improved by an analysis of the Fisher matrix, as shown in~\cite{Smith:2006ud}. 

For a separable shape, the Fisher matrix can also be expressed by a sum over contributions of different $r$ sampling points that arise when numerically evaluating the bispectrum integral in eq. (\ref{eq_reducedbispectrum1}). The total Fisher matrix is then given as a sum $F = \sum_{i,j=1}^{N_{fact}} F_{ij}$, where $F_{ij}$ is the Fisher matrix element between the sampling points $i$ and $j$ and $N_{fact}$ is the number of sampling points. The Fisher matrix elements are then given by
\begin{align}
\label{eq_fisher5} 
F_{ij}= \frac{1}{6} \sum_{l_1l_2l_3} \frac{(2l_1+1) (2l_2+1)(2l_3+1)}{4\pi} \nonumber\\
\times \left( \begin{array}{ccc}
l_1 & l_2 & l_3 \\
0 & 0 & 0 \end{array} \right)^2
 \frac{b^i_{l_1l_2l_3} b^j_{l_1l_2l_3}}{C_{l_1}C_{l_2}C_{l_3}}.
\end{align}
We now explicitly consider the sine term of the linear model (the cosine term is analogous), where the contribution of a distance $r_i$ is given by
\begin{equation}
\label{eq_fisher3} 
b^i_{l_1l_2l_3} =  (\Delta r_i) r_i^2~ \big[-X_{l_1}(r_i) X_{l_2}(r_i) X_{l_3}(r_i) + \left[ X_{l_1}(r_i) Y_{l_2}(r_i) Y_{l_3}(r_i) + \textrm{2 perm.}\right] \big].
\end{equation}

\begin{figure}
\centering
\resizebox{0.8\hsize}{!}{
\includegraphics{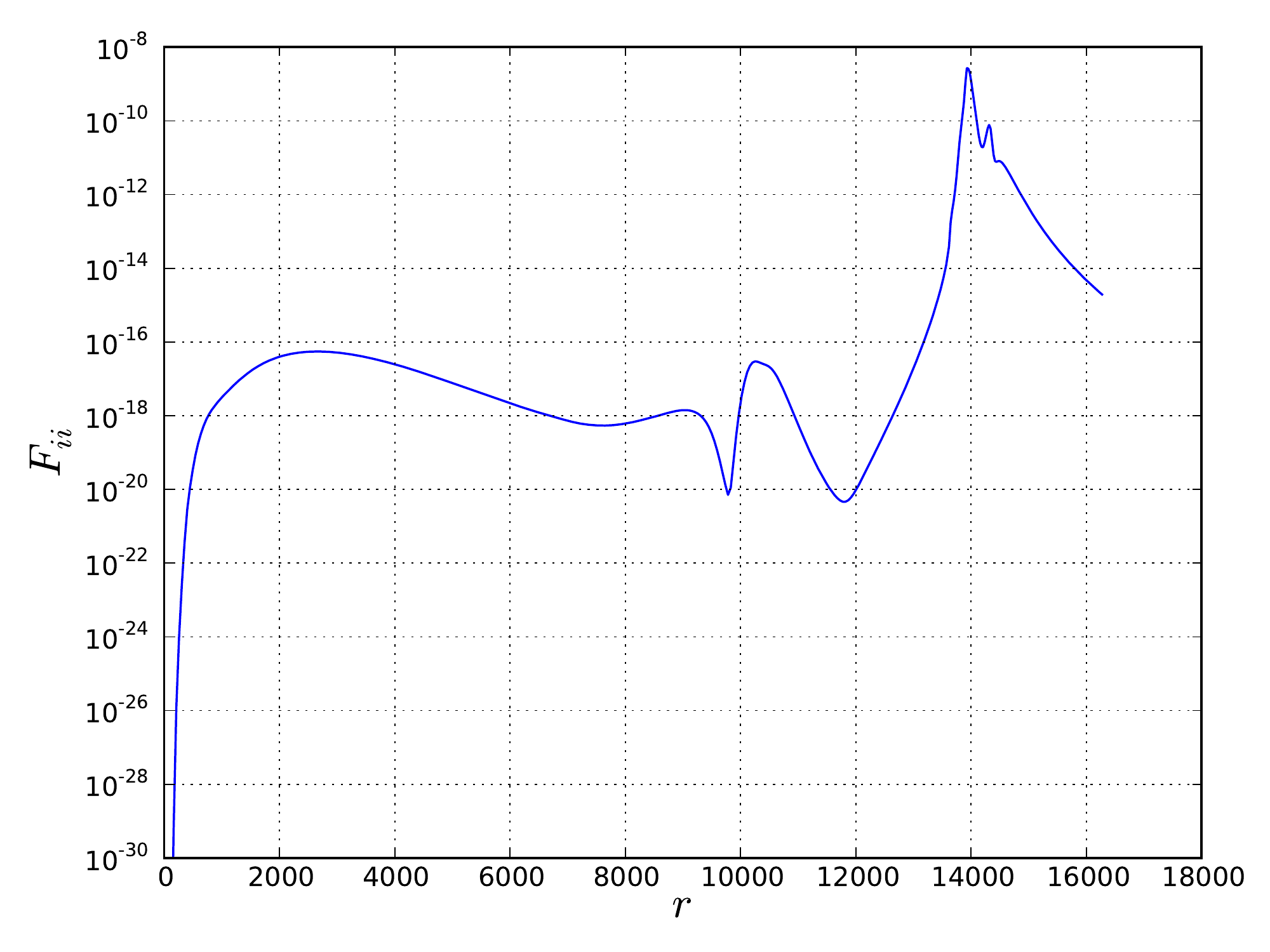}
}
\resizebox{0.8\hsize}{!}{
\includegraphics{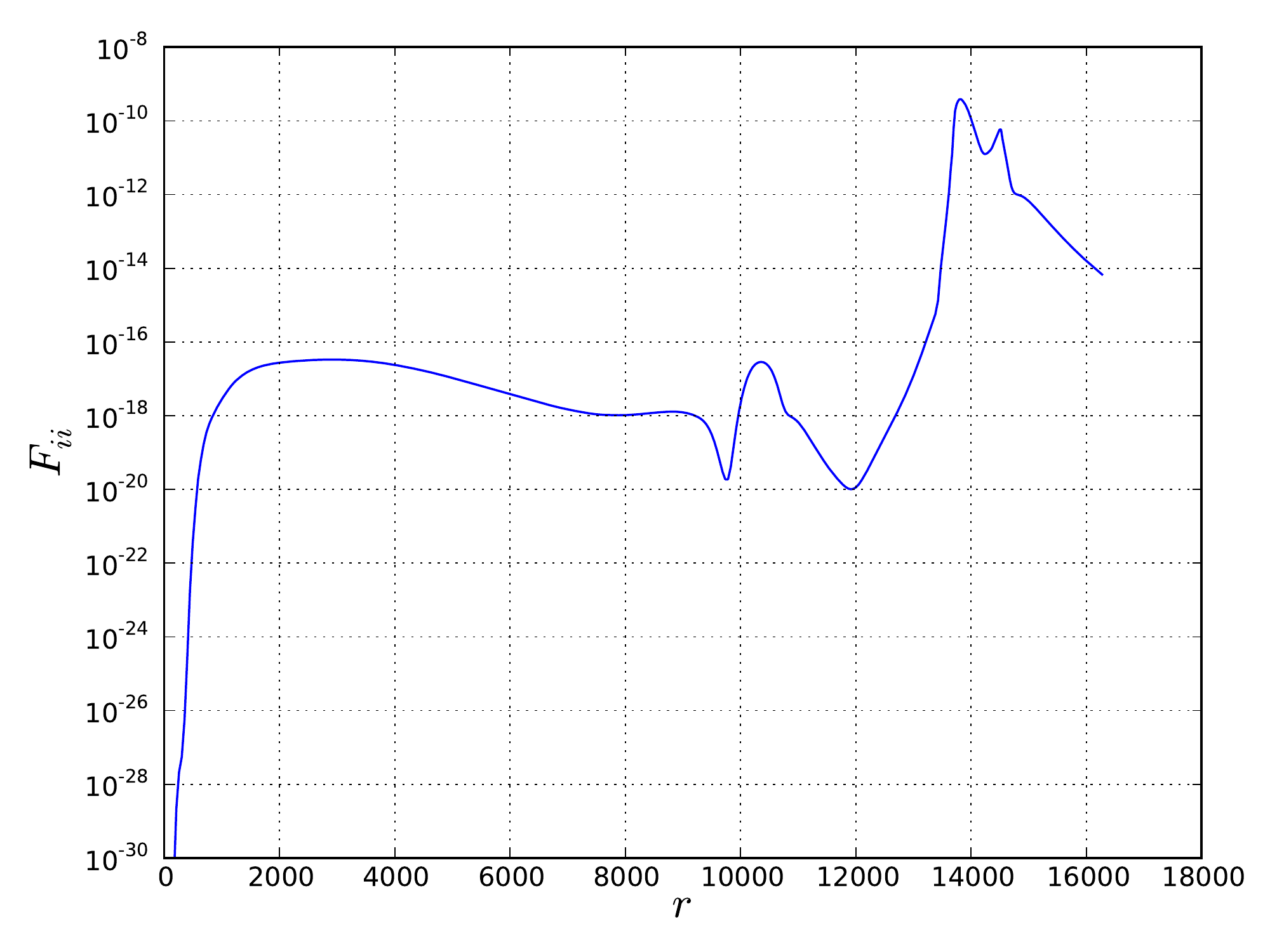}
}
\caption{$F_{ii}$ as a function of conformal distance $r_i$. Top: $k_c=0.01$, bottom:
  $k_c=0.005$.} 
\label{fig:Fii}
\end{figure}

To give an impression of the bispectrum contribution of different distances $r_i$, we plot the diagonal elements $F_{ii}$ in figure \ref{fig:Fii}. The plots show the contribution of recombination ($r=14000$ Mpc), reionisation ($r\simeq 10500$ Mpc), and ISW ($r>5000$ Mpc). As expected, the dominant contribution comes from the time around recombination, and one can get a good approximation to the integral by sampling only a window around recombination. This is particularly useful to quickly scan a parameter space, in the present case the oscillation frequency, $k_c$. 

In~\cite{Smith:2006ud}, it was shown that one can go further and optimise the $r$ sampling points to find a quadrature with surprisingly few sampling points that give an almost identical estimator. Their algorithm constructs a new bispectrum $B'$ from the original bispectrum $B$ by choosing a subsample of points and weighting them so that the Fisher distance between the two is minimised:
\begin{equation}
\label{eq_smith1} 
F(B,B') = \frac{1}{6} \sum_{l_1l_2l_3} \frac{(B_{l_1l_2l_3} - B'_{l_1l_2l_3})^2}{C_{l_1}C_{l_2}C_{l_3}}.
\end{equation}
This means that bispectrum values with a small signal-to-noise are allowed to be very different. Using this algorithm, as an example, we obtain an approximate bispectrum $B'$ consisting of 30 sample points leading to a separability between $B$ and $B'$ of $0.1 \sigma$ assuming $k_c=0.01$ and $f_{NL}=1000$.

The most computationally demanding task in the estimation pipeline remains the calculation of the Fisher matrix in equation \ref{eq_fisher5}. It was also shown in~\cite{Smith:2006ud} that one can factorise this equation by inserting the integral representation of the Wigner symbol 
\begin{equation}
\label{eq_wignersymbol} 
\left( \begin{array}{ccc}
l_1 & l_2 & l_3 \\
0 & 0 & 0 \end{array} \right)^2
= \frac{1}{2} \int_{-1}^{1} dz P_{l_1}(z) P_{l_2}(z) P_{l_3}(z).
\end{equation}
The integral over $z$ can be computed efficiently by Gauss Legendre integration. However, even with this expression, the calculation of $F_{ij}$ needs many CPU hours depending on the chosen initial quadrature point number. If one does not want to calculate an optimised quadrature at every frequency points, but only wants to know the normalisation $F(k_c)$ of the estimators, one can calculate this normalisation on a much wider frequency spacing and interpolate in between. This can be seen from the frequency dependent Fisher matrix in figure \ref{fig:fishermatrix}, where the diagonal elements vary slowly.

\section{A position space interpretation of the KSW estimator for oscillations}

In appendix A, we show that the three-point function of the feature model in position space peaks for configurations $\hat{n}_1,\hat{n}_2 ,\hat{n}_3$ that lie on a circumcircle of radius $\sin(\theta) = 1/(2k_0\eta)$, where $2k_0 = \frac{3}{2\pi} k_c$ in the convention of eq. (\ref{eq_oscispectrum1}). This suggests searching for bispectrum oscillations in real space by convoluting the CMB map with a ring kernel of varying radius. 

\begin{figure}
\resizebox{\hsize}{!}{
\includegraphics{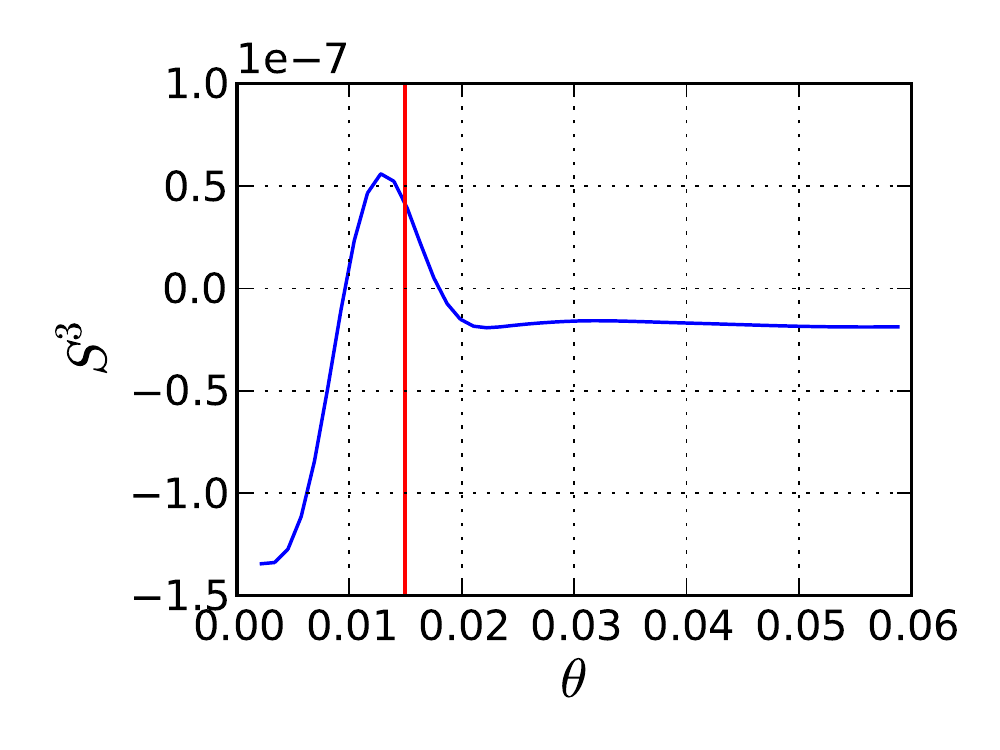}
\includegraphics{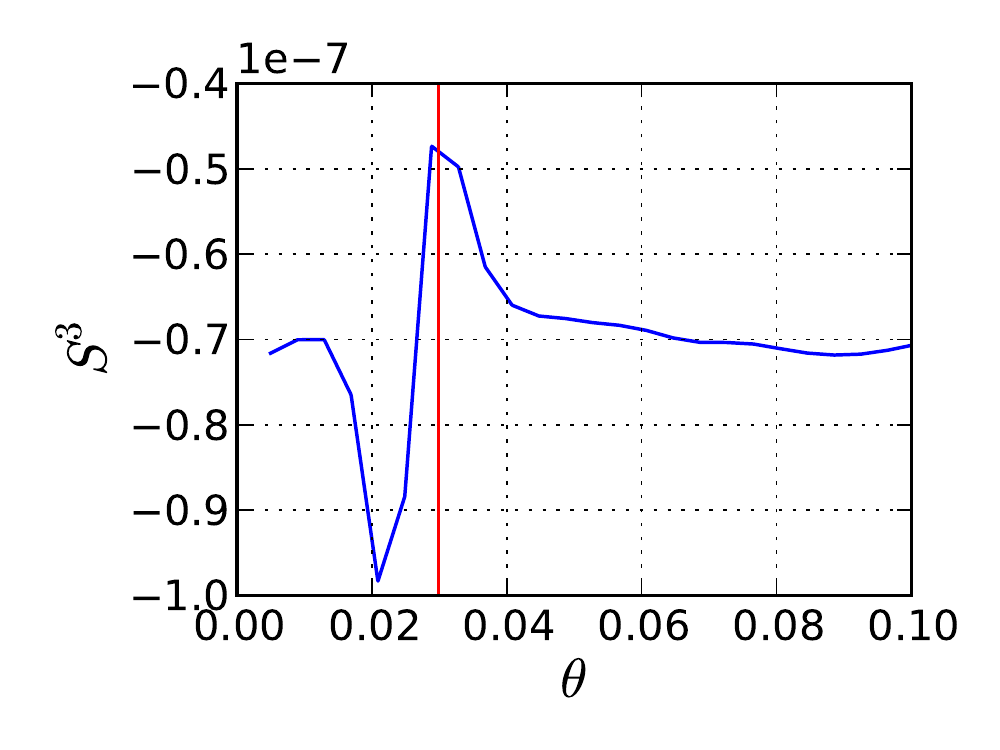}
}
\resizebox{\hsize}{!}{
\includegraphics[width=0.02\linewidth]{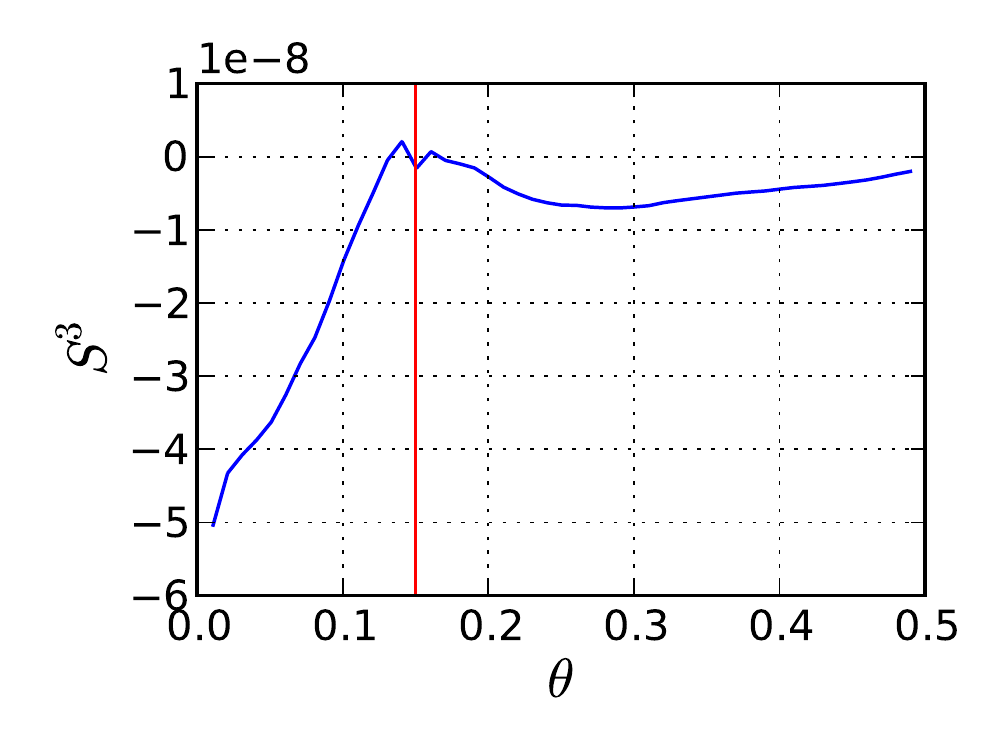}
}
\caption{Ring estimate $\mathcal{E}^{ring}_{NG} - \mathcal{E}^{ring}_{G}$. Top left: $k_c=0.01$. Top right:
  $k_c=0.005$. Bottom: $k_c=0.001$. The kernel width in all plots was $\Delta \theta = 0.005$, $f_{NL}$ was 10-20 times the optimal Fisher error. The red line shows the expected position of the maximum.} 
\label{fig:ringestimate}
\end{figure}

For a radially symmetric kernel, the convolution can be done efficiently in harmonic space as $s_{lm}=K_l r_{lm}$, where the kernel is given by the Legendre transformation,
\begin{equation}
\label{eq_kernel2} 
K_l = 2 \pi \int_{-1}^{1}  K(z) P_l(z) dz,
\end{equation}
and $K(z)$ is a narrow window function in $z=\cos(\theta)$. The estimate $\mathcal{E}^{ring}(z)$ is given by the sum over the pixels of the cube of the convoluted map,
\begin{equation}
\label{eq_ringestimate} 
\mathcal{E}^{ring} (z_0) = \int d\Omega \left[ \sum_{lm} K_l a_{lm} Y_{lm} \right]^3.
\end{equation}
Figure \ref{fig:ringestimate} shows an example of this estimator for a map that was simulated with the algorithm of the preceding section. The plots show $\mathcal{E}^{ring}_{NG} - \mathcal{E}^{ring}_{G}$, which means that the estimate from the Gaussian map was subtracted. The red line shows the predicted position of the maximum. We note that this maximum would be harder to locate in real data where the Gaussian contribution cannot simply be subtracted.

\begin{figure}
\resizebox{\hsize}{!}{
\includegraphics{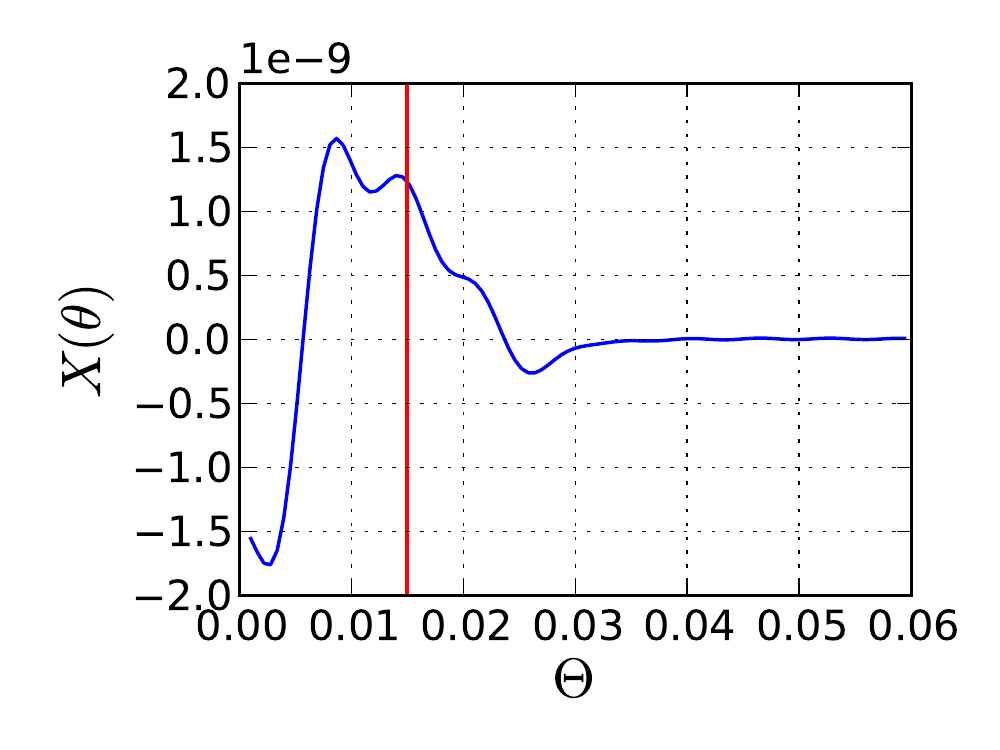}
\includegraphics{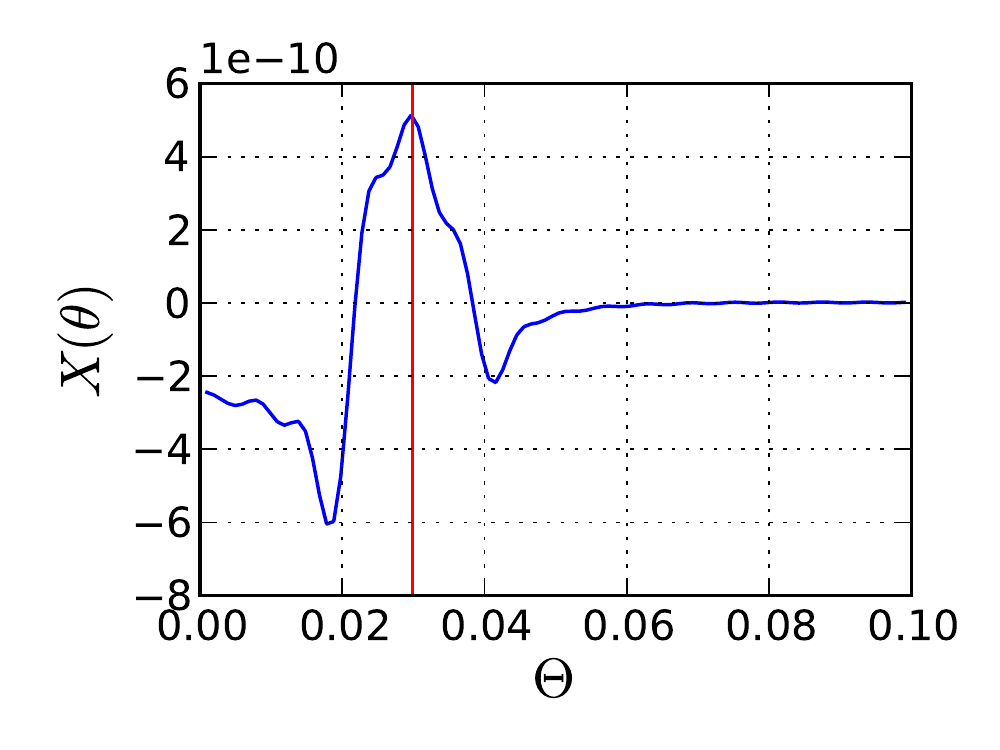}
}
\resizebox{\hsize}{!}{
\includegraphics[width=0.02\linewidth]{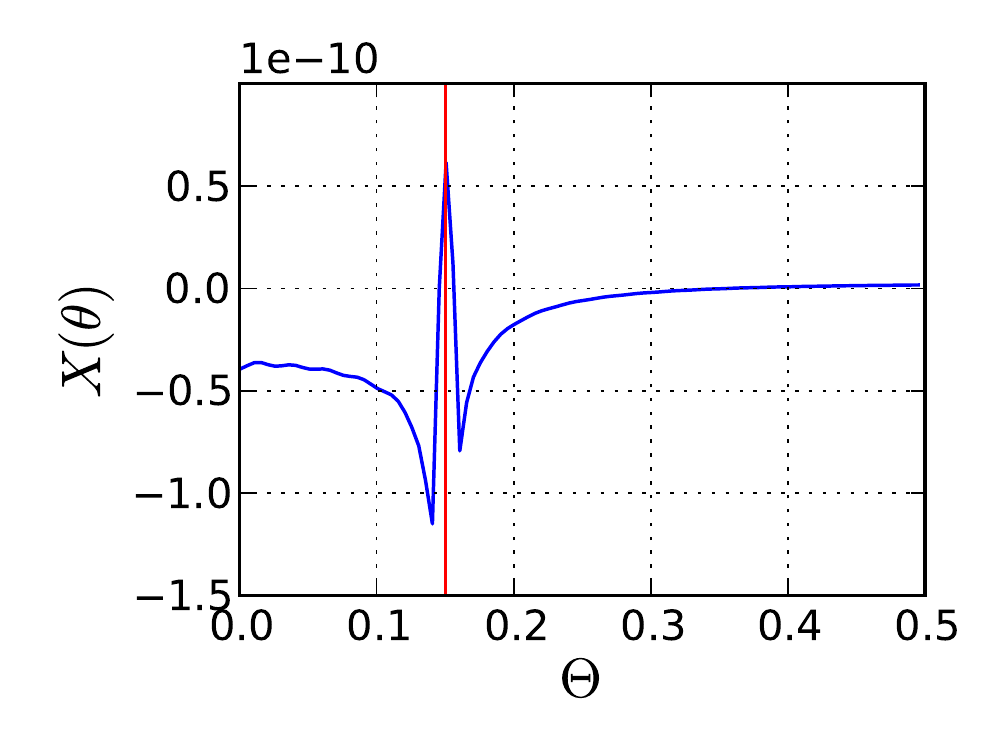}
}
\caption{KSW kernel $X(\theta)$ at decoupling radius $r_{rec}$. Top left: $k_c=0.01$. Top right:
  $k_c=0.005$. Bottom: $k_c=0.001$. The red line is the predicted
  maximum of the kernel. } 
\label{fig:kswkernel_realspace}
\end{figure}

It is interesting to compare the ``intuitive'' ring kernel $K(\theta)$ with the KSW kernel that is known to give optimal results. The KSW estimator is of form,
\begin{equation}
\label{eq_kernel3} 
\mathcal{E}^{KSW}(a) \propto  \int r^2 dr \int d\Omega \left[ \sum_{lm} \frac{X_l a_{lm}}{C_l} Y_{lm} \right]^3 + ... .
\end{equation}
The largest contribution to this integral comes from decoupling at $r_{rec}$. We plot the KSW kernel function at decoupling $X_{rec}(\theta)$ in figure \ref{fig:kswkernel_realspace}. It shows the maximum at the expected position and has roughly the expected window shape. 

\section{Conclusion}

In this paper, we have presented and extensively studied the KSW estimator for linear bispectrum oscillations. The main motivation for this approach is that the oscillating bispectrum shapes are difficult to represent with a modal expansion and, thus, have not yet been constrained at high oscillation frequency. We have provided the equations for estimation and map making and validated them with Monte Carlo simulations. Unlike many of the well know bispectum shapes, oscillations have two free parameters in addition to the common amplitude parameter $f_{NL}$. We have developed the methodology to estimate and constrain the oscillation phase $\phi$ and the frequency $k_c$. Our work will therefore allow to explore a parameter space that was not previously accessible for a theoretically well-motivated bispectrum shape. Finally, we have found an interesting position space interpretation of the KSW estimator for oscillations, based on an approximate evaluation of the corresponding three-point function.

\begin{acknowledgements}
The authors acknowledge support from NSF Grant NSF AST 09-08693 ARRA. BDW acknowledges funding from an ANR Chaire d'Excellence (ANR-10-CEXC-004-01), the UPMC Chaire Internationale in Theoretical Cosmology, and NSF grants AST-0908 902 and AST-0708849. This work made in the ILP LABEX (under reference ANR-10-LABX-63) was supported by French state funds managed by the ANR within the Investissements d'Avenir programme under reference ANR-11-IDEX-0004-02. MM acknowledges funding by Centre National d'Etudes Spatiales (CNES). The authors would like to thank Daan Meerburg for useful discussions and comments. 
\end{acknowledgements}

\bibliographystyle{aa} 
\bibliography{mybib} 

\begin{appendix}
\section{Angular correlation function of the feature model} \label{App:AppendixA}

In this appendix, we show that the linear feature model bispectrum peaks in real space for a special class of three-point function configurations. A similar analysis was presented in~\cite{Adshead:2011jq}. The corresponding calculation for logarithmic oscillations can be found in~\cite{Jackson:2013mka}.

In the approximation of instantaneous CMB decoupling at time $\eta$, the CMB temperature perturbation is given in terms of the potential $\phi$ as~\cite{Bashinsky:2000uh},
\begin{equation}
\label{dirFourier}
 \frac{\Delta T({\hat {\bf n}})}{T} = \int \frac{ d^3 {\bf k}}{(2 \pi)^3} \ \varphi_{\bf k}(0^-) T_{\rm rad}(k) e^{-i \eta {\bf k} \cdot {\hat {\bf n}}} .
\end{equation} 
The transfer function $T_{\rm rad}(k)$ is generally a complicated function of scale and cosmological parameters.  For simplicity of an analytic answer, which accounts for finite resolution, we take it to be $ T_{\rm rad}(k) \approx e^{-k^2/k_D^2} $. 

The position-space primordial bispectrum is given by
\begin{align}
\langle  \frac{\Delta T({\hat {\bf n}}_1)}{T}&  \frac{\Delta T({\hat {\bf n}}_2)}{T}  \frac{\Delta T({\hat {\bf n}}_3)}{T} \rangle = \nonumber \\ 
 &\int \prod_{i=1}^3 \frac{ d^3 {\bf k}_i}{(2 \pi)^3} T_{\rm rad}(k_i) \exp \left( - i \eta  {\bf k}_i  \cdot {\hat {\bf n}}_i \right) \langle \varphi_{{\bf k}_1} \varphi_{{\bf k}_2} \varphi_{{\bf k}_3} \rangle,
\end{align}
where the correlation is of the form,
\[ \langle \varphi_{{\bf k}_1} \varphi_{{\bf k}_2} \varphi_{{\bf k}_3} \rangle  \equiv B_\varphi( {\bf k}_1, {\bf k}_2, {\bf k}_3)  (2 \pi)^3 \delta^3 \left( \sum_{i=1}^3 {\bf k}_i  \right) . \]
Since ${\bf k}_1+{\bf k}_2+{\bf k}_3=0$, the $k$-space correlations are categorised by the triangle formed by the ${\bf k}_i$. 

We now examine this integral for the linear model, 
\[ B_\varphi = \frac{B_0}{ (k_1 k_2 k_3)^2} \sin \left(  \frac{k_1 + k_2 + k_3}{2k_0} \right) .\]
The delta function that couples the momenta can be written as
$$ (2 \pi)^3 \delta^3 \left( {\bf k}_1 +  {\bf k}_2 +  {\bf k}_3 \right) = \eta^3 \int d^3 {\bf w} \ e^{ -i \eta {\bf w} \cdot ( {\bf k}_1 +  {\bf k}_2 +  {\bf k}_3  ) },$$
where the factor of $\eta$ has been included for future convenience. Writing the sine as exponentials and performing the integral over angles, we obtain
\begin{align*}
\left( \frac{\Delta T}{T} \right)^3_{\rm osc} &= \frac{B_0 \eta^3}{2i} \int d^3 {\bf w} \sum_\pm  \prod_{i=1}^3 \int \frac{d^3 {\bf k}_i}{(2 \pi)^3 k_i^2} \\
&\times \pm e^{ \pm i k_i/2k_0}  e^{-i \eta {\bf k}_i \cdot ({\bf w} +   {\hat {\bf n}_i}) } e^{ - k_i^2/ k_D^2} \\
& =  \frac{B_0 \eta^3}{2i} \int d^3 {\bf w} \sum_\pm \prod_{i=1}^3 \frac{ \pm 1}{(2 \pi)^2 i  \eta | {\bf w} + {\hat {\bf n}_i}|} \\
& \times \int_0^\infty \frac{dk_i}{k_i} \ e^{\pm i k_i/2k_0} \left( e^{-i k_i \eta | {\bf w} +  {\hat {\bf n}_i}|} - e^{ik_i \eta | {\bf w} +  {\hat {\bf n}_i}|} \right) e^{-k_i^2/k_D^2}.
\end{align*}
Defining the dimensionless parameter $x_i~\equiv~k\eta$, the momentum integral is 
\begin{align*}
2 \int_0^\infty \frac{dx_i}{x_i}  &\Big( \cos \big[ (\eta k_0)^{-1} +| {\bf w} +  {\hat {\bf n}_i}| \big] x_i \nonumber \\
&- \cos \left[ (\eta k_0)^{-1} - | {\bf w} +  {\hat {\bf n}_i}| \right]x_i  \Big) e^{ - 2x_i^2/\eta^2 k_D^2}.
\end{align*}
This integral can be evaluated exactly in terms of a hypergeometric function, but there is a simplifying limit we can take.  The low-momentum, long-distance approximation allows
\[ \cos \left[  (k_0 \eta)^{-1} \pm | {\bf w} +  {\hat {\bf n}_i}|  \right] x \approx e^{ -\left[  (k_0 \eta)^{-1} \pm | {\bf w} +  {\hat {\bf n}_i}|  \right]^2 x^2/2} . \]
The integral then has the simple analytic solution 
\begin{align*}
\left( \frac{\Delta T}{T} \right)^3_{\rm osc}  \approx& \frac{B_0 \eta^3}{2i} \int d^3 {\bf w} \ \prod_{i=1}^3  \frac{ 1}{(2 \pi)^2 i  \eta | {\bf w} + {\hat {\bf n}_i}|} \nonumber \\
&\ln \left( \frac{  \left[ (2k_0 \eta)^{-1} + | {\bf w} +  {\hat {\bf n}_i}| \right]^2 + 2(\eta k_D)^{-2}}{ \left[ (2k_0 \eta)^{-1} -  | {\bf w} +  {\hat {\bf n}_i}| \right]^2 + 2(\eta k_D)^{-2}} \right).
\end{align*}
In the $\eta k_D \gg 1$ limit, this maximally peaks when all three products peak near
\[ ( 2 \eta k_0)^{-1} - |{\bf w}_0 + {\hat {\bf n}_i}| = 0 , \hspace{0.5in} i=1, 2, 3. \]
Squaring then subtracting, we obtain
\[ {\bf w}_0 \cdot {\bf N}_{ij} = 0, \hspace{0.5in} {\bf N}_{ij} \equiv  {\hat {\bf n}_i} -  {\hat {\bf n}_j}, \hspace{0.4in} i \neq j. \]
We can take the three vectors ${\bf N}_{12}$, ${\bf N}_{23}$, and ${\bf N}_{31}$ and arrange them in the ${\hat {\bf x}}-{\hat {\bf y}}$ plane, so that
\begin{align*}
{\hat {\bf n}_i} &= \sin \theta \cos \Theta_i {\hat {\bf x}} + \sin \theta \sin \Theta_i {\hat {\bf y}} + \cos \theta {\hat {\bf z}}, \\
{\bf w}_0 &= (\pm \rho -\cos \theta ) {\hat {\bf z}},
\end{align*}
where 
\[ \rho = \sqrt{ (2 \eta k_0)^{-2} - \sin^2 \theta }. \]
We now have the value of ${\bf w}$ for which the three factors are in resonance for a given configuration $({\bf n}_1,{\bf n}_2,{\bf n}_3)$. The value of the factor at the resonance point is largest when $| {\bf w} +  {\hat {\bf n}_i}|$ is at its minimum, which is the case for $\rho=0$. We conclude that the three-point function peaks for triangle configurations $({\bf n}_1,{\bf n}_2,{\bf n}_3)$ that lie on a circle with radius given by $\sin(\theta) = \frac{1}{2 k_0 \eta}$.

\end{appendix}

\end{document}